\documentclass[published]{JHEP3}

\JHEP{10(2004)070}
\received{July 27, 2004}
\revised{October 8, 2004}
\accepted{October 28, 2004}

\skip\footins = 1\bigskipamount plus 2pt minus 4pt

\renewcommand{\>}{\rangle}
\newcommand{\beq}{\begin{equation}}
\newcommand{\eeq}{\end{equation}}
\newcommand{\beqn}{\begin{eqnarray}}
\newcommand{\eeqn}{\end{eqnarray}}
\newcommand{\nn}{\nonumber}
\newcommand{\wt}{\widetilde}
\newcommand{\half}{\frac{1}{2}}
\newcommand{\U}{\mathop{\rm {}U}\nolimits}
\newcommand{\SU}{\mathop{\rm SU}\nolimits}

\title{Chirally improving Wilson fermions\\
2. Four-quark operators}

\author{Roberto Frezzotti\\
INFN, Sezione di Milano\\
Dipartimento di Fisica, Universit\`a di Milano
``Bicocca''\\
Piazza della Scienza 3,  20126 Milano, Italy\\
E-mail: \email{Roberto.Frezzotti@mib.infn.it}}

\author{Giancarlo Rossi\\
Dipartimento di Fisica, Universit\`a di  Roma ``Tor Vergata''\\
INFN, Sezione di Roma ``Tor Vergata''\\
Via della Ricerca Scientifica,  00133 Roma, Italy, and\\
NIC/DESY Zeuthen, Platanenallee 6, D-15738 Zeuthen, 
Germany\\
E-mail: \email{rossig@roma2.infn.it}}

\abstract{In this paper we discuss how the peculiar properties of
twisted mass lattice QCD at maximal twist can be employed to set up a
consistent computational scheme in which, despite the explicit
breaking of chiral symmetry induced by the presence of the Wilson
and mass terms in the action, it is possible to completely bypass
the problem of wrong chirality and parity mixings in the
computation of the CP-conserving matrix elements of the $\Delta
S=1,2$ effective weak hamiltonian, and at the same time have a
positive determinant for pairs of non-degenerate quarks, as well as 
absence of O($a$) discretization effects in on-shell quantities 
with no need of improving lattice action and operators.}

\begin{document}

\section{Introduction}
\label{sec:INTRO}

In this paper we want to give an explicit set of rules that,
within the computational scheme offered by the maximally twisted
lattice regularization of QCD~\cite{TMLQCD,FR1}, allows to extract
from simulation data the matrix elements of the CP-conserving part
of the $\Delta S=1,2$ effective weak hamiltonian
(${\cal{H}}_{\rm{eff}}^{\Delta S=1}$ and
${\cal{H}}_{\rm{eff}}^{\Delta S=2}$) between pseudo-scalar mesons
with no contamination from mixing with operators of wrong
chirality or parity, nor O($a$) discretization errors. In
contrast, a rather complicated mixing pattern has to be
considered in order to construct multiplicatively renormalizable
chirally covariant operators if, instead, the standard Wilson
fermion regularization is employed~\cite{BMMRT}.

To achieve these remarkable results we will have to treat on
different footing internal (sea) and boundary (valence) quark
lines. Non-degenerate sea quarks are introduced in pairs and
regularized as described in~\cite{FR2}. In this way one gets a
real, positive determinant with no vanishing eigenvalues for
non-zero values of the quark mass, thus also solving the well
known problems with the spectrum of the standard Dirac-Wilson
operator. We immediately note that, through quark loops, all
lattice quantities at finite $a$ will obviously depend on the set
of Wilson parameters, $\{r_p\}$, associated to all the dynamical
pairs of quarks ($p=\ell{\rm{(ight)}}, h{\rm{(eavy)}},\ldots$)
present in the action.

As for valence quarks, a precise choice of their flavour
structure and regularization scheme has to be made. In this paper
we will discuss in detail the case in which valence quarks are
introduced \`a la \"Osterwalder and Seiler~\cite{OS,FR1}, 
but one could equally well imagine to employ fermions obeying the
Ginsparg-Wilson (GW) relation~\cite{GW} for this purpose~\cite{BaRuSh}.
In this case O($a$) improvement and no wrong chirality mixing is also
achieved, though at the expenses of a fairly larger computational
burden compared to what is necessary for the inversion of the
standard or twisted Wilson-Dirac fermion operator. It should be
immediately observed, however, that the effort to invert, say, the
overlap~\cite{OVER,LU} operator is not expected to be larger
than that of stochastically evaluating the twisted determinant.

In this paper we want to stick to Wilson-like fermions. We
will show that one can still attain the ambitious goals mentioned
above within the twisted mass regularization of QCD, with a
judicious choice of the structure of the regularized valence quark
action. More precisely, we propose to describe each 
valence flavour, $f$, with an action of the type 
\beqn
S_{\rm{OS}}^{(\pi/2)}[q_f,\bar q_f,U]&=& a^4 \sum_x\,\bar q_f(x)
\Big{[}\gamma\cdot\widetilde\nabla+e^{-i\omega\gamma_5}
W_{\rm{cr}}(r_f)+m_f\Big{]}\Big{|}_{\omega=\pi/2}q_f(x)\qquad
\nn\\
&=&a^4 \sum_x\,\bar q_f(x) \Big{[}\gamma\cdot\widetilde\nabla
-i\gamma_5 W_{\rm{cr}}(r_f)+m_f\Big{]}q_f(x)\, ,
\label{VALQACT}
\eeqn
where
\beqn
\gamma\cdot\widetilde\nabla&\equiv&\frac{1}{2}\sum_\mu\gamma_\mu
(\nabla^\star_\mu+\nabla_\mu)\, ,
\label{WDONDD}\\
W_{\rm{cr}}(r_f)&\equiv&-a\frac{r_f}{2}\sum_\mu\nabla^\star_\mu\nabla_\mu+
M_{\rm{cr}}(r_f)\, , 
\label{WDONDM}
\eeqn 
with $r_f$ the Wilson parameter for the flavour $f$ and
$M_{\rm{cr}}(r_f)$ the corresponding critical quark mass. As remarked
above, in the full theory $M_{\rm{cr}}(r_f)$ also depends on the set
of sea quark parameters $\{r_p\}$, though for short we have not
indicated this dependence.\footnote{Unless differently stated, we
employ here the notations of ref.~\cite{FR1}. In particular, the
ubiquitous dependence on the bare gauge coupling, $g_0^2$, will always
be omitted.} We will specify below more carefully what is to be meant
by ``critical mass'' in the present context. We wrote the valence
quark action in the so-called ``physical basis''~\cite{FR1}, where
$m_f$ is taken to be real (and positive).

The key feature we will exploit in the following is the freedom to
regularize each flavour with an appropriately chosen
value (of the sign) of the $r$ parameter in front of the corresponding
Wilson term.

One might wonder at this point whether the scheme we are proposing is
a consistent one, precisely in view of the fact that we are
regularizing quarks in different manners according to whether they are
sea or valence quarks. It is not difficult to convince oneself that,
provided renormalized sea and valence quark masses of the same flavour
are made to coincide, this is indeed so. The idea of the argument is
to introduce compensating lattice ghost fields (\`a la
Morel~\cite{MOR}) to exactly cancel the fermion determinant of the
valence quarks and exploit the symmetries of the extended (local)
theory~\cite{SHSH} to prove that the Green functions of operators
constructed in terms of only ordinary fields have the expected
renormalization properties as the lattice spacing, $a$, is sent to
zero (see appendix~\ref{sec:APPA} for some detail).

We will show that, by replicating some of the valence flavours (and
accompanying ghosts) and regularizing them with Wilson terms of
suitably chosen signs, the resulting lattice action turns out to have
a sufficiently large symmetry so as to imply the cancellation of all
the unwanted mixings in the lattice computation of the matrix elements
of the CP-conserving ${\Delta S=1}$ and ${\Delta S=2}$ effective weak
hamiltonian.  It is important to remark that all these modifications
of the lattice action can be made without spoiling the ``magic"
improvement properties of maximally twisted LQCD~\cite{FR1}.

A preliminary study of the scaling properties of maximally twisted
LQCD in the quenched approximation has been carried out in
ref.~\cite{KIAC}, where measurements of the pion decay constant and
the mass of the vector particle have been performed in the interval
$\beta=5.85\div 6.2$.  Data extrapolate very smoothly to the continuum
limit with residual discretization errors consistent with the expected
O($a^2$) behaviour.

Concluding, we wish to cite the nice paper of ref.~\cite{PSV} in which
an interesting step in the direction of alleviating the
renormalization problems of the ${\cal{H}}_{\rm{eff}}^{\Delta S=1,2}$
operators within the framework of twisted mass LQCD (tm-LQCD) has been
taken. The authors show that, with an appropriate choice of the
twisting angles of the various flavours, in the computation of
$K\to\pi$ transitions it is possible to remain with at most a linearly
divergent counterterm, which in some partially quenched approximation
can be even reduced to a finite one.

The plan of the paper is as follows. In section~\ref{sec:VQ} we
discuss how valence quarks must be introduced and regularized in order
not to loose O($a$) improvement and at the same time be ready to get
cancellation of wrong chirality and parity mixings in the relevant
physical amplitudes. In section~\ref{sec:BK} we present the strategy
for the calculation of $B_K$, showing how in practice one can get rid
of O($a$) discretization errors and completely dispose of the
contributions potentially coming from the mixing of the bare operator
${\cal{H}}_{\rm{eff}}^{\Delta S=2}$ with operators of unwanted
chirality and parity. In sections~\ref{sec:DIUM1} and~\ref{sec:DIUM2}
we extend this procedure to the calculation of the $K\to\pi$ and
$K\to\pi\pi$ matrix elements of the CP-conserving
${\cal{H}}_{\rm{eff}}^{\Delta S=1}$ operator. A few concluding remarks
can be found in section~\ref{sec:CONCL}. Technical details are
relegated in some appendices.

\section{Sea and valence quarks}
\label{sec:VQ}

In this paper we will consider QCD with four active quark
flavours, a situation which represents a rather realistic
framework for the computation of the hadronic matrix elements of
the effective weak hamiltonian.
For the reasons explained in the Introduction we regularize
in a different way sea and valence quarks and thus
we take the lattice action to be of the form 
\beq 
S = S_g[U]
+\sum_{p=\ell,h} S_{\rm tm}^{(\pi/2)}[\psi_p,\bar{\psi}_p,U]+
\sum_f \Big{[} S_{\rm OS}^{(\pi/2)}[q_f,\bar{q}_f,U] +
S_{gh}^{(\pi/2)}[\phi_f,U] \Big{]} \,, 
\label{FULACT}
\eeq 
where $U$ are the gauge link variables and $S_g[U]$ stands for any
sensible discretization of the pure gauge action. We briefly comment
on the three ``fermionic" bits of this action.
\begin{itemize}
\item Sea quarks ($p=\ell,h$) are arranged in
non-degenerate pairs: $\psi_\ell$ with components $u_{\rm sea},d_{\rm
sea}$ (${\it\ell}$ight doublet) and $\psi_h$ with components $s_{\rm
sea},c_{\rm sea}$ (${\it h}$eavy doublet). As proposed in
ref.~\cite{FR2}, they will be described by the maximally twisted
action\footnote{With respect to ref.~\cite{FR2} we have inverted the
sign in front of the mass splitting term.}
\beq
S_{\rm tm}^{(\pi/2)}[\psi_p,\bar{\psi}_p,U] = a^4 \sum_x\,\bar
\psi_p(x) \Big{[}\gamma\cdot\widetilde\nabla -i\gamma_5\tau_1
W_{\rm{cr}}(r_p)+m_{p} - \epsilon_{p}\tau_3
\Big{]}\psi_p(x)\label{SEAQACT} 
\eeq 
with $\tau_1$ and $\tau_3$ acting in flavour space.  Integration over
the sea quark fields leaves behind a real positive (for
$|\epsilon_{p}| < |m_{p}|$, $p=\ell,h$) determinant (see
ref.~\cite{FR2} and section~\ref{sec:SQ}).

\item Valence quarks are denoted by $q_f$. Each
valence quark will be described by the OS lattice action, $S_{\rm
OS}^{(\pi/2)}[q_f,\bar{q}_f,U]$, specified in eq.~(\ref{VALQACT}). 
Since it will turn out to be necessary to introduce ``replicas'' of
the usual $u,d,s,c$ quark fields, we immediately notice that the
range of variability of the index $f$ will cover the more extended
set of values $u$, $u'$, $u''$, \dots, $d$, $d'$, $d''$, \dots,
$s$, $s'$, $s''$, \dots, $c$, $c'$, $c''$, \dots.

\item In order to have a local representation of
correlators with insertion of valence quarks, each $q_f$
will be introduced with an accompanying ghost, $\phi_f$,
which has the only purpose of exactly canceling the corresponding
valence determinant~\cite{MOR}. The action of each ghost reads
\beq
S_{gh}^{(\pi/2)}[\phi_f,U] = a^4 \sum_x\, \phi^\dagger_f(x)
\, {\rm sign}(m_f) \Big{[}\gamma\cdot\widetilde\nabla -i\gamma_5
W_{\rm{cr}}(r_f)+m_f\Big{]}\phi_f(x)\, .\label{VALGACT}
\eeq
The (complex) field $\phi_f$ is an euclidean \emph{commuting} spinor in the
fundamental representation of the gauge group with precisely the 
same value of the mass parameter, $m_f$, and the same lattice 
regularization as the quark field, $q_f$, it corresponds to.
Since we will never need to consider correlators with insertions
of ghost fields, nor graded field transformations (which turn
valence ghosts into valence quarks and viceversa), we do not need
to introduce the sophisticated and powerful formalism developed
in~\cite{SHSH}.

The factor ``${\rm sign}\,(m_f)$" in the ghost action~(\ref{VALGACT})
is inserted in order to guarantee the convergence of the gaussian
integral over the $\phi_f$ field in case $m_f<0$~\cite{SHSH}.  Indeed,
with the purpose of discussing renormalizability and O($a$)
improvement of the correlation functions computed with the
action~(\ref{FULACT}), in the following we will introduce spurionic
symmetries that involve the inversion of the sign of $m_f$. Thus the
case of negative $m_f$ arises, even if one starts with
$m_f>0$.\footnote{Notice that the factor ${\rm sign}(m_f)$ does not
introduce any spurious non-analytic behaviour of the ghost determinant
in $m_f$, as its presence only amounts to rescaling the determinant by
an overall field-independent factor, which in any case drops out from
(connected) correlators.}
\end{itemize}

As it is formulated, the lattice model with action~(\ref{FULACT}) can
be viewed as an euclidean statistical theory where the approach of
correlation functions to the continuum limit can be discussed on the
basis of standard power counting and symmetry arguments.  In
particular it turns out that renormalized correlation functions admit
a well defined continuum limit provided, besides suitably rescaling
the bare gauge coupling and setting the critical mass parameters of
sea and valence quarks (and corresponding ghosts) to the appropriate
values, for each flavour renormalized masses of sea and valence quarks
(and ghosts) are given the same numerical value. Actually a continuum
limit exists even if renormalized sea and valence quark masses of the
same flavour are given different values: however, we will not consider
the possibility of partial quenching in this paper.

Energy levels and operator matrix elements of QCD with four quark
flavours can be extracted from correlation functions of the
model~(\ref{FULACT}) with neither ghost nor sea quark fields.  As for
correlators with sea quark fields, they are also perfectly well
defined and are needed, for instance, to compute the renormalization
constants $Z_P$ and $Z_S$ (see eq.~(\ref{SEAMQRE}) below and
appendix~\ref{sec:APPA}).

\subsection{Symmetries and renormalization}
\label{sec:WTIS}

Since the lattice regularizations of sea and valence quarks are not
identical, implementing the equality of sea and valence
renormalized quark masses requires the knowledge of the relation
between bare and renormalized masses. Such a relation follows, as
usual, from the analysis of the symmetries of the model described
by the action~(\ref{FULACT}). The result of this analysis, which is 
presented in appendix~\ref{sec:APPA}, can be summarized as follows.
\begin{itemize}
\item A unique (dimensionless) lattice function,
$f_{\rm{cr}}(r_1;r_2,r_3)$, exists in terms of which all critical
masses can be expressed.  Besides the coupling constant (which, as we
said, we do not show explicitly), $f_{\rm{cr}}$ depends on the Wilson
parameter, $r_1$, of the quark whose critical mass is being considered
and on the Wilson parameters of sea quarks. The former --- which
appears only in the ``external" lines of the diagrams relevant for the
study of quark mass renormalization --- can be $r_f$, or either
$r_\ell$ or $r_h$, depending on whether we are dealing with valence or
sea quark masses. The further dependence upon the Wilson parameters
$r_2 \equiv r_\ell$ and $r_3 \equiv r_h$ is the one that comes from
``internal" loops.

In the following for short the dependence on sea quark
$r$-coefficients ($r_\ell$ and $r_h$) will always be understood.
Thus the relevant formulae for the critical masses will be written
in the form 
\beqn
&{\rm{sea}} \qquad
&aM_{\rm{cr}}(r_p)=f_{\rm{cr}}(r_p;r_\ell,r_h)\, ,
\quad p=\ell,h\, ,  
\label{CRMS} \\
&{\rm{valence}} \qquad
&aM_{\rm{cr}}(r_f)=f_{\rm{cr}}(r_f;r_\ell,r_h) \, ,\quad
\hbox{all }f'{\rm{s}}\, .  \label{CRMV}
\eeqn
It may be worth noting that also for valence quarks the value of the
critical mass is determined by the requirement that $\bar
q_f\gamma_5W_{\rm cr}q_f$ (eq.~(\ref{WDONDM})) is a truly dimension
five (irrelevant) operator.
\item The renormalized quark mass parameters are related to
their bare counterparts by the relations
\begin{eqnarray}
\hat{m}^{\pm}_{p}&=&Z_{P}^{-1}(r_p)m_{p}
\pm Z_{S}^{-1}(r_p)\,\epsilon_{p} \, ,
\qquad p=\ell,h \, ,
\label{SEAMQRE} \\
\hat{m}_f &=& Z_m\,(r_f) m_f \, , \qquad \hbox{all }f{\rm 's}\, ,
\label{VALMQRE}
\end{eqnarray}
where $\pm$ refers to the high and low mass components of each sea quark 
pair. Also in the above renormalization constants the dependence on the
Wilson parameters that enter only through internal loops will be
understood in the following. In other words, we will write
$Z_I(r)$ as a shorthand for ${Z}_I(r; r_\ell, r_h)$,
$I=m,P,S,\dots$.
\item $Z_{P}(r_p)$ and $Z_{S}(r_p)$ are the renormalization
constants of the (non-singlet) sea quark bilinears $\bar\psi_p
\gamma_5 \tau_{2,3} \psi_p$ and $\bar\psi_p \tau_{2,3} \psi_p$,
respectively. They can be determined by requiring that the chiral
Ward-Takahashi identities (WTI's) of the model~(\ref{FULACT}) with
insertions of sea quark field operators have the form expected in
the corresponding continuum target theory.
\item $Z_m(r_f)$ is the inverse renormalization constant of the
operator $\bar q_f q_f$. The renormalized expression of the latter, in
the model we are considering, is $Z_m^{-1}(r_f)[\bar q_f q_f+m_f
c_{\bar{q}q}(r_f,am_f) a^{-2}]$, where the dimensionless coefficient
$c_{\bar{q}q}$ is even in $r_f$ and~$am_f$.\footnote{These properties of
$c_{\bar{q}q}$ follow from the spurionic invariances ${\cal
R}_{5f}^{\rm sp}$ (eq.~(\ref{R5SPVAL})) and ${\cal P}_5 \times (M \to
- M)$ (eqs.~(\ref{P5}) and~(\ref{R_M_def})).}  $Z_m(r_f)$ is even in
$r_f$, as it follows, for instance, from the argument given below
eq.~(\ref{RCONT}), and obeys the relation
\beq
Z_m(r)\equiv{Z}_m(r;r_\ell,r_h) = {Z}_P^{-1}(r;r_\ell,r_h)\, .
\label{ZMZP}
\eeq
\item Conservation of each of the $N$ valence flavour quantum
numbers is ensured by the invariance of~(\ref{FULACT}) under the
$\U(1)$ transformations ($f=1,\ldots,N$)
\beq 
{\cal I}_f:\left
\{\begin{array}{ll} q_f(x) \rightarrow e^{i\theta_f} q_f(x) \, ,
\quad\; &
\bar{q}_f(x) \rightarrow \bar{q}_f(x) e^{-i\theta_f}\\
q_g(x) \rightarrow q_g(x) \, , \quad &\bar{q}_g(x) \rightarrow
\bar{q}_g(x) \quad\quad {\rm if} \  g\neq f
\end{array}\right . \label{FIXFLAROT}
\eeq
with $\theta_f$ a real parameter. Larger vector valence flavour
symmetries of course exist if several valence quark species have
identical Wilson parameters (and bare masses).
\item The flavour singlet axial current $J_{5\mu} = \sum_{f=1}^{N}
\bar{q}_f \gamma_\mu\gamma_5 q_f$, made out of exclusively valence
quarks, exhibits the correct anomaly, proportional to $N$~\cite{SEST}.
\end{itemize}
Generalizing the arguments given in ref.~\cite{FR1}, one can prove
that $f_{\rm{cr}}(r_1;r_2,r_3)$ is odd in $r_1$.  For instance, if we
take $r_1=r_f$, proving that $f_{\rm{cr}}$ is odd in its first
argument is equivalent to prove that $M_{\rm{cr}}(r_f)$ is odd under
$r_f \to -r_f$.  This result follows (see appendix~\ref{sec:APPA} for
details) from the observation that under the changes of integration
variables
\beq
{\cal{R}}_{5f} : \left \{\begin{array}{rl} q_f(x) &
\rightarrow q'_f(x)=\gamma_5 q_f(x) \\
\bar{q}_f(x) &\rightarrow\bar{q}'_f(x) =-\bar{q}_f(x) \gamma_5 \\
\phi_f(x) & \rightarrow \phi'_f(x) = \gamma_5 \phi_f(x)
\end{array}\right.
\label{R5}
\eeq
the valence quark (plus ghost) action is invariant if we also invert the
sign of $r_f$ and $m_f$~\cite{FR1}.

An analogous argument applies if we were to consider the critical
mass of sea quarks. In this case one has to perform the change of
variables
\beq
{\cal{R}}_{5p} : \left \{\begin{array}{rl}
\psi_p(x) & \rightarrow \psi'_p(x)=\gamma_5 \psi_p(x)  \\
\bar{\psi}_p(x) &\rightarrow\bar{\psi}'_p(x) =-\bar{\psi}_p(x) \gamma_5
\end{array}\right.  \label{R5SEA}
\eeq
and exploit the fact that for each value of $p$ the fermionic determinant
is an even function of the corresponding Wilson parameter, $r_p$
(see section~\ref{sec:SQ}).

In the following for any $r\in\{r_\ell,r_h;\dots,r_f,\dots\}$
it is understood that $W_{\rm cr}(r)$ is of the
form~(\ref{WDONDM}) with $M_{\rm cr}(r)$ the appropriate odd function 
of $r$ specified in eq.~(\ref{CRMS}) or eq.~(\ref{CRMV}). 
We will thus label expectation values by the relevant 
Wilson parameters $\{r_\ell,r_h;\dots,r_f,\dots\}$ and quark
masses $\{m_\ell,\epsilon_\ell, m_h,\epsilon_h;\dots,m_f,\dots\}$,
for which we introduce the compact notation
\beq 
M \equiv \{ m_f \, , {\rm all}\; f{\rm's ;} \, m_{p}\, ,
\epsilon_{p}\, , p=\ell,h \} \, , \qquad R \equiv \{ r_f \, , {\rm
all}\; f{\rm's ;} \, r_p\, , p=\ell,h \} \, . 
\label{R_M_def} 
\eeq

\subsection{Sea quark determinant}
\label{sec:SQ}

Integration over sea quark fields gives rise to
the product of determinants 
\beqn
{\rm{D}}&=&\prod_{p=\ell,h} \, {\rm{D}}_{p}\, ,
\label{DET}\\
{\rm{D}}_{p}&=&{\rm Det}\Big{[} \gamma\cdot\widetilde\nabla
-i\gamma_5\tau_1 W_{\rm{cr}}(r_p)+m_{p} - \epsilon_{p}\tau_3
\Big{]} \, . 
\label{SEADET} 
\eeqn 
As proved in~\cite{FR2}, each factor in ${\rm{D}}$ is even in $m_{p}$
and $\epsilon_{p}$. From this result one can show that ${\rm{D}}_{p}$
is also even in $r_p$. To prove this property we follow again the
strategy developed in ref.~\cite{FR1}. By performing in the functional
integral that defines ${\rm{D}}_{p}$ the change of
variables~(\ref{R5SEA}), we get the identity
\beq
{\rm Det}\Big{[}\gamma\cdot\widetilde\nabla
-i\gamma_5\tau_1 W_{\rm{cr}}(r_p)+m_{p} - \epsilon_{p}\tau_3\Big{]} =
{\rm Det}\Big{[}\gamma\cdot\widetilde\nabla +i\gamma_5\tau_1
W_{\rm{cr}}(r_p)-m_{p} + \epsilon_{p}\tau_3\Big{]}\, .
\label{SYMDET}
\eeq
Since ${\rm{D}}_{p}$ is even in $m_{p}$ and $\epsilon_{p}$, while
$W_{\rm{cr}}(r_p)$ is odd in $r_p$, we get the thesis.  It then
follows that all the correlators with insertion of operators made out
of only valence quark and gluon fields, and thus all the quantities
derived from them, are even in $r_p$, $m_{p}$ and $\epsilon_{p}$,
$p=\ell,h$. In particular such parity properties imply that
$f_{\rm{cr}}(r_1;r_2,r_3)$ is an even function of its last two
arguments ($r_{2,3}=r_{p}, p=\ell,h$).

\subsection{Renormalization constants and O($a$) improvement}
\label{sec:RCI}

The symmetries of the action~(\ref{FULACT}) relevant for analyzing the 
$R$-dependence of renormalization constants and leading cut-off effects are
\begin{itemize}
\item the spurionic ${\cal R}_{5f}^{\rm sp}$ transformation acting
on each flavour $f$ separately
\beq
{\cal R}_{5f}^{\rm sp} \equiv {\cal R}_{5f}\times (r_f\rightarrow
-r_f)\times (m_f\rightarrow -m_f)\, ,
\label{R5SPVAL}
\eeq
with ${\cal{R}}_{5f}$ given by
eq.~(\ref{R5}); \item the spurionic ${\cal R}_{5p}^{\rm sp}$
transformation acting on the $p=\ell$ and $p=h$ sea quarks
separately,
\beq
{\cal R}_{5p}^{\rm sp} \equiv {\cal R}_{5p}
\times (r_p \rightarrow -r_p) \times (m_{p}\rightarrow
-m_{p})\times (\epsilon_{p}\rightarrow -\epsilon_{p}) \, ,
\label{R5SPSEA} 
\eeq
with ${\cal{R}}_{5p}$ given by
eq.~(\ref{R5SEA}); \item the overall transformation
acting on all types of fields
\beq 
{\cal D}_d
\times \prod_f {\cal R}_{5f} \times \prod_{p=l,h} {\cal R}_{5p}
\eeq
with
\begin{equation}
{\cal{D}}_d : \left \{\begin{array}{rclrcll}
U_\mu(x)&\rightarrow& U_\mu^\dagger(-x-a\hat\mu) \\
\psi_p(x)&\rightarrow& e^{3i\pi/2} \psi_p(-x) \, ,\qquad
&\bar{\psi}_p(x)&\rightarrow& e^{3i\pi/2} \bar{\psi}_p(-x)
\, , \quad&p=\ell,h  \\
q_f(x)&\rightarrow& e^{3i\pi/2} q_f(-x) \, ,
&\bar{q}_f(x)&\rightarrow& e^{3i\pi/2} \bar{q}_f(-x) \, ,
\quad&\hbox{all } f{\rm's}  \\
\phi_f(x)&\rightarrow& e^{3i\pi/2} \phi_f(-x)\, , &&&\quad&\hbox{all } f{\rm's} \, .
\end{array}\right . 
\label{Dd}
\end{equation}
\end{itemize}
We also note the existence of the spurionic symmetry obtained as
the product of all invariances of the type~(\ref{R5SPVAL})
and~(\ref{R5SPSEA}), namely 
\beq 
{\cal{R}}^{\rm{L}}\times (M
\rightarrow -M)\equiv\prod_f {\cal R}_{5f} \times \prod_{p=\ell,h}
{\cal R}_{5p} \times (R \rightarrow -R) \times (M \rightarrow -M)
\, . 
\label{R5SPTOT} 
\eeq
Since the formal continuum action we are aiming at, of
which~(\ref{FULACT}) is the regularized version, will have to be 
invariant under the (discrete, non-anomalous) spurionic
transformation
\beq
{\cal{R}}^{\rm{cont}}\times (M \rightarrow
-M)\equiv\prod_f {\cal R}_{5f} \times \prod_{p=l,h} {\cal R}_{5p}
\times (M \rightarrow -M) \, , 
\label{RCONT}
\eeq
it follows that in any renormalization scheme, besides $Z_g$, the
renormalization constants $Z_P$, $Z_S$ and $Z_m$, appearing in
eqs.~(\ref{SEAMQRE}) and~(\ref{VALMQRE}), and those of all the local
operators whose bare expression has definite transformation properties
under ${\cal{R}}^{\rm{L}}$, will be even functions of $R$. This is, in
fact, necessary for the renormalized operators to behave under the
latter transformation as the corresponding bare ones. Furthermore,
recalling that the fermionic determinant D$_p$ (eq.~(\ref{SEADET})) is
even in $r_p$, one concludes that all the above renormalization
constants will be separately even in (or trivially independent of)
$r_p$, $p=\ell,h$ and $r_f$, all $f$'s.

\subsubsection{{\rm O}($a$) improvement via Wilson averaging}
\label{sec:IWA}

By a direct generalization of the argument given in
ref.~\cite{FR1}, the above symmetries and remarks imply that
$WA$'s of expectation values of multiplicatively renormalizable
(m.r.), multi-local, gauge invariant operators (with no ghost
fields), $O(x_1,x_2,\dots,x_n)$, are O($a$) improved, i.e.\ 
\begin{eqnarray}
\langle O(x_1,\dots ,x_n) \rangle\Big{|}_{(M)}^{WA}&\equiv&
\frac{1}{2}\Big{[}\langle O(x_1,\dots ,x_n) \rangle\Big{|}_{(R,M)}+
\langle O(x_1,\dots ,x_n) \rangle\Big{|}_{(-R,M)}\Big{]} \qquad
\nonumber \\
&=& \zeta^{O}_{O}(R)
\langle O(x_1,\dots ,x_n) \rangle\Big{|}^{\rm{cont}}_{(M)}+{\rm{O}}(a^2)\, ,
\label{WILAV}
\end{eqnarray}
where it is understood that the space-time arguments $x_1$, \dots
$x_n$ are at non-zero relative distance in physical units. The
multiplicative factor $\zeta^{O}_{O}(R)$ is a coefficient (even
under $R \to - R$) whose presence is necessary to match the
lattice to the corresponding continuum expectation value.

Analogous results can be derived from eq.~(\ref{WILAV}) for energies and
operator matrix elements, namely
\begin{eqnarray}
E_{h,n}({\bf k};R,M)+ E_{h,n}({\bf k};-R,M) &=& 
2 E^{\rm cont}_{h,n}({\bf k},M)+{\rm O}(a^2)\, , 
\label{ENWA}\\[5pt]
\langle h,n,{\bf k}
|{\cal O}|h',n',{\bf k}'\rangle |_{(R,M)} +&&
\nonumber\\[-3pt]
+\langle h,n,{\bf k} |{\cal O}|h',n',{\bf k}'\rangle
|_{(-R,M)}&=& 2  \zeta^{{\cal O}}_{{\cal O}}(R)
\langle h,n,{\bf k}|{\cal O}|h',n',{\bf k}'\rangle\!\Big{|}^{\rm{cont}}_{(M)}
+{\rm{O}}(a^2) \, ,\qquad
\label{MEWA}
\end{eqnarray}
where $E_{h,n}({\bf k};R,M)$ is the energy of $|h,n,{\bf
k}\rangle|_{(R,M)}$, the $n$-th eigenstate of the lattice quantum
hamiltonian (which we assume to exist) with three-momentum ${\bf
k}$ and unbroken quantum numbers $h$, while ${\cal O}$ denotes an
arbitrary m.r.\ and gauge invariant local operator.

\subsubsection{{\rm O}($a$) improved results from a single simulation}
\label{sec:IRSS}

The action~(\ref{FULACT}) admits the further spurionic symmetry
\beq
{\cal P} \times (R \rightarrow -R) \, , \label{PSP} 
\eeq
where ${\cal{P}}$ is the parity operation. The latter, with the
definition $x_P=(-{\bf{x}},t)$, has in the present context the
form
\begin{equation}
{\cal{P}}:\left \{\begin{array}{rclrcll}
 U_0(x)&\rightarrow& U_0(x_P)\, ,\qquad&
 U_k(x)&\rightarrow& U_k^{\dagger}(x_P-a\hat{k})\, ,
\quad& k=1,2,3
\\
\psi_p(x)&\rightarrow& \gamma_0 \psi_p(x_P) \, ,&
\bar{\psi}_p(x)
&\rightarrow&\bar{\psi}_p(x_P)\gamma_0 \, , \quad& p=\ell,h 
\\
q_f(x)&\rightarrow& \gamma_0 q_f(x_P) \, , &
\bar{q}_f(x)
&\rightarrow&\bar{q}_f(x_P)\gamma_0 \, , \quad&\hbox{all } f{\rm's} 
\\
\phi_f(x)&\rightarrow& \gamma_0 \phi_f(x_P) \, , &&&\quad& \hbox{all } f{\rm's} \,.
\end{array}\right . \label{PAROP}
\end{equation}
The spurionic symmetry~(\ref{PSP}) is particularly important, as
its existence entails the possibility of associating a definite
parity label to m.r.\ local operators and to eigenstates of the
lattice hamiltonian.\footnote{The assignment of a parity label to
eigenstates of the hamiltonian can be done in close analogy with
the procedure outlined in sections~5.2 and~5.3 of ref.~\cite{FR1},
where the case of twisted mass QCD with one pair of mass
degenerate quarks was discussed.} By arguments similar to those
given in ref.~\cite{FR1} one can prove the validity of the
symmetry relations
\beqn
E_{h,n}({\bf k};-R,M) \,
&=& E_{h,n}({\bf -k};R,M) \, ,
\label{ENIMPR}\\
\langle h,n,{\bf k} |{\cal O}|h',n',{\bf
k}'\rangle \!\Big{|}_{(-R,M)} & =& \eta_{hn} \eta_{\cal O}
\eta_{h'n'} \langle h,n,{\bf -k} |{\cal O}|h',n',{\bf
-k}'\rangle\!\Big{|}_{(R,M)} \, , \qquad
\label{MEIMPR} 
\eeqn
where $\eta_{\cal O}$, $\eta_{hn}$ and $\eta_{h'n'}$ denote the parity
labels of the m.r.\ operator ${\cal O}$ and the external states,
respectively. Inserting eqs.~(\ref{ENIMPR}) and~(\ref{MEIMPR})
in~(\ref{ENWA}) and~(\ref{MEWA}), one ends up with the formulae
\begin{eqnarray}
E_{h,n}({\bf k};R,M)+ E_{h,n}(-{\bf k};R,M) &=& 
2 E^{\rm cont}_{h,n}({\bf k},M)+{\rm O}(a^2)\, , 
\label{ENWAS}\\[5pt]
\langle h,n,{\bf k} |{\cal O}|h',n',{\bf k}'\rangle
|_{(R,M)} + 
\nonumber \\[-3pt]
+\eta_{hn} \eta_{\cal O} \eta_{h'n'}
\langle h,n,-{\bf k} |{\cal O}|h',n',-{\bf k}'\rangle
|_{(R,M)}\!&=& 
2  \zeta^{{\cal O}}_{{\cal O}}(R)
\langle h,n,{\bf k}|{\cal O}|h',n',{\bf k}'\rangle\!\Big{|}^{\rm{cont}}_{(M)}
\!+{\rm{O}}(a^2)  ,\qquad
\label{MEWAS}
\end{eqnarray}
which show that the $WA$'s~(\ref{ENWA}) and~(\ref{MEWA}) can be evaluated
in terms of correlators and derived quantities obtained from a single
simulation with Wilson parameters $R$.

\subsection{Some conclusive observations}
\label{sec:OSS}

It is worth observing that in the lattice model~(\ref{FULACT}) the
fermionic determinant~(\ref{DET}) introduces neither lattice
artifacts linear in $a$ nor parity breaking effects.
This can be seen by considering expectation values of pure gauge
(multi-local, gauge invariant) operators, 
${O}_{pg}$. Such expectation values depend on the sea quark
Wilson parameters, $r_\ell$ and $r_h$, and are even functions of
them, because ${\rm D}={\rm D}_\ell\,{\rm D}_h$ is even, but are
obviously independent of the valence quark Wilson parameters. We
thus have
\beq
\label{R_even_VEV}
\langle {O}_{pg}(x_1,\dots
,x_n) \rangle\Big{|}_{(R,M)} = \langle {O}_{pg}(x_1,\dots
,x_n) \rangle\Big{|}_{(-R,M)} \, , 
\eeq
from which, comparing with
the $WA$ formula~(\ref{WILAV}) written for
${O}\to{O}_{pg}$, we conclude that the expectation
values of pure gauge operators trivially coincide with their
Wilson average and, hence, are automatically free of O($a$) cut-off effects.
Moreover, if the operator ${O}_{pg}$ has a definite parity,
$\eta_{{O}_{pg}} = \pm 1$, i.e.\ it satisfies the equation
\beq
\label{P_def_OP}
\langle {O}_{pg}(x_1,\dots ,x_n)
\rangle\Big{|}_{(R,M)} = \eta_{{O}_{pg}} \langle
{O}_{pg}(x_{1P},\dots ,x_{nP}) \rangle\Big{|}_{(-R,M)}
\, , 
\eeq
then for its expectation value, owing to
eq.~(\ref{R_even_VEV}), one gets \beq \label{P_def_OPVEV} \langle
{O}_{pg}(x_1,\dots ,x_n) \rangle\Big{|}_{(R,M)} =
\eta_{{O}_{pg}} \langle {O}_{pg}(x_{1P},\dots ,x_{nP})
\rangle\Big{|}_{(R,M)} \, , 
\eeq
which is precisely what one would
expect in a parity-invariant theory. In other words, all parity
breaking effects in correlators with insertions of valence quark
operators (and not containing sea quarks or ghosts) come from the
breaking of parity induced by the valence OS quarks themselves,
and reduce to O($a^2$) in Wilson averaged quantities.

The basic reason behind the validity of these properties is that, as a
functional of the set of link variables, $\{U\}$, the (maximally
twisted) fermion determinant, ${\rm{D}}={\rm{D}}[\{U\}]$, is invariant
under the replacement $\{U\}\to \{U_{P}\}$, where $\{U_{P}\}$ is the
parity transformed set of link variables. This result can be proved by
performing in the functional integral that defines the determinant,
${\rm{D}}$, the change of sea quark integration variables induced by
the transformation in the second line of eq.~(\ref{PAROP}) and
recalling that {\rm D} is even in $r_p,p=\ell,h$.

All these observations may be of some importance in practice, because
they suggest the interesting possibility of employing GW-type fermions
to regularize valence quarks and maximally twisted Wilson fermions to
describe sea quarks.\footnote{A very similar proposal, with ordinary
(rather than maximally twisted) Wilson fermions playing the role of
sea quarks, has been made in ref.~\cite{BaRuSh}.} In this hybrid
framework, whose computational cost does not look prohibitively large
on the scale of today's available computer resources, O($a$)
improvement is automatic. Furthermore, the nice chiral properties of
GW fermions are sufficient to ensure absence of all wrong chirality
mixings in correlators with insertions of only valence quark and gluon
fields.

This last statement might not be obvious. Actually in the next three
sections we will show that this remarkable result can be achieved even
if valence quarks are regularized \`a la OS, provided some clever
adjustment of the lattice regularization of (four-flavour) QCD and of
its computational scheme is made.

\section{The $B_K$ parameter}
\label{sec:BK}

The popular way to discuss the issue of CP violation in the Standard
Model is through the study of the compatibility of experimental data
with the constraints imposed by the unitarity of the CKM matrix
(unitarity triangle) involving the third generation of quark
flavours.\footnote{Good reviews on the subject can be found in
ref.~\cite{BURAS_REV}. For recent summaries of lattice results see the
papers quoted in ref.~\cite{BKLAT}.} One of the key quantity entering this
analysis is $\epsilon_K$, which represents the amount of indirect
CP-violation in $K\to\pi\pi$ decay.  Exploiting its precise
experimental value to constrain the non-trivial vertex of the
unitarity triangle requires the knowledge of the parameter $B_K$.

In the formal continuum euclidean QCD with four quark flavours
($u$, $d$, $s$ and $c$), renormalized at the scale $\mu$, the kaon
$B$-parameter, $B_K(\mu)$, is defined by the formula 
\beq
\langle
\bar{K}^0 | \hat{\cal O}_{VV+AA}^{\Delta S=2} (\mu) | K^0\rangle =
{16 \over 3} M_K^2 F_K^2  B_K(\mu) \, . \label{BKDEF} 
\eeq
$B_K$
can be extracted from the correlator ($x_0\!>\!0$, $y_0\!<\!0$)
\beq C_{K{\cal O}K}^{({\rm qcd}4)}(x,y) = \langle (\bar{d}\gamma_5
s)(x) \widehat{{\cal O}}_{VV+AA}^{\Delta S=2}(0) (\bar{d}\gamma_5
s)(y) \rangle \, ,
 \label{CORRBK}
\eeq
where the bare expression of the renormalized operator
$\widehat{{\cal O}}_{VV+AA}^{\Delta S=2}$ is
\beq {\cal
O}_{VV+AA}^{\Delta S=2} = (\bar{s} \gamma_\mu d) (\bar{s}
\gamma_\mu d) + (\bar{s} \gamma_\mu\gamma_5 d) (\bar{s}
\gamma_\mu\gamma_5 d) \label{OPBKDEF}
\eeq
and the superscript $({\rm qcd}4)$ denotes QCD with four quark species.

We want to show that it is possible to set up a lattice
regularization based on (twisted) Wilson fermions, where $B_K$ can be 
computed in terms of a four-quark operator which does not mix with any other
operator of the theory.

In order to explain our strategy it is convenient to
first consider at a formal level an auxiliary gauge model
where, besides the four lightest sea quark flavours ($u_{\rm
sea}$, $d_{\rm sea}$, $s_{\rm sea}$ and $c_{\rm sea}$), each of
the $d$ and $s$ valence quark species is duplicated. The valence
quarks $u$, $d$, $d'$, $s$, $s'$ and $c$ are introduced together with the
corresponding ghosts, required to cancel the associated valence determinant. 
We will denote the correlators of this model by the superscript
$(4s6v)$. It is not difficult to convince oneself that in this
model the correlator
\beq 
C_{K'{\cal Q}K}^{(4s6v)}(x,y) = \langle
(\bar{d}'\gamma_5 s')(x) \, 2 {\cal Q}_{VV+AA}^{\Delta S=2}(0)
(\bar{d}\gamma_5 s)(y) \rangle \, , \label{COR2BK} 
\eeq
where\footnote{What we really mean with the superscript ${\Delta
S=2}$ in ${\cal Q}_{VV+AA}^{\Delta S=2}$ is ${\Delta s+\Delta
s'=2}$.}
\begin{eqnarray}
 {\cal Q}_{VV+AA}^{\Delta S=2} &=&
(\bar{s} \gamma_\mu d) (\bar{s}' \gamma_\mu d')
+ (\bar{s} \gamma_\mu\gamma_5 d) (\bar{s}' \gamma_\mu\gamma_5 d') +
\nonumber \\
&& + (\bar{s} \gamma_\mu d') (\bar{s}' \gamma_\mu d)
+ (\bar{s} \gamma_\mu\gamma_5 d') (\bar{s}' \gamma_\mu\gamma_5 d) \, ,
\label{OPBKCONT}
\end{eqnarray}
contains the same physical information as the ``target''
correlator $C_{K{\cal O}K}^{({\rm qcd}4)}(x,y)$ in four-flavour QCD.

To see this we simply have to imagine to have regularized
the two models in the same way, for instance, by using GW fermions 
for all quark fields. If the bare masses of the quarks of the same
flavour are given the same value, i.e.\ if one sets 
\beq
\begin{array}[b]{rclrcl}
m_u^{({\rm qcd4})} &=& m_u^{(4s6v)} = m_{u_{\rm
sea}}^{(4s6v)} \, , \qquad &
m_d^{({\rm qcd4})} &=& m_d^{(4s6v)} =
m_{d'}^{(4s6v)} = m_{d_{\rm sea}}^{(4s6v)} \, ,
\\[3pt]
m_c^{({\rm qcd4})} &=& m_c^{(4s6v)} = m_{c_{\rm
sea}}^{(4s6v)} \, , \qquad& 
m_s^{({\rm qcd4})} &=& m_s^{(4s6v)} =
m_{s'}^{(4s6v)} = m_{s_{\rm sea}}^{(4s6v)} \, , 
\end{array}
\eeq
then by an
elementary application of the Wick theorem one checks that the
equality
\beq
C_{K'{\cal Q}K}^{(4s6v)}(x,y) = C_{K{\cal
O}K}^{({\rm qcd}4)}(x,y) \, \label{WITHCONT} 
\eeq
holds (sea quark loops trivially contribute the same to the two sides
of the relation) for any value of the lattice spacing. This result
implies that the two operators, ${\cal O}_{VV+AA}^{\Delta S=2}$ in the
$({\rm qcd4})$ model and ${\cal Q}_{VV+AA}^{\Delta S=2}$ in the
$(4s6v)$ model, can be taken to have the same renormalization
constants. The same is also true for the renormalization constants of
the operator $\bar{d} \gamma_5 s$ in the $({\rm qcd4})$ model and the
operators $\bar{d} \gamma_5 s$ and $\bar{d}' \gamma_5 s'$ in the
$(4s6v)$ model. Moreover, it is obvious that the renormalized masses
of the quarks of the same flavour are (can be chosen to be) equal in
the two models. With this natural choice of renormalization
conditions, the relation~(\ref{WITHCONT}) holds for the corresponding
renormalized correlators and carries over to the continuum limit.

The strategy we propose for the evaluation of $B_K$ is based on
eq.~(\ref{WITHCONT}) and consists in considering the correlator
$C_{K'{\cal Q}K}^{(4s6v)}(x,y)$ in the UV regularization provided
by the maximally twisted action~(\ref{FULACT}). The reason for
selecting this UV regularization is that, with a careful choice of
the Wilson parameters of valence quarks, ${\cal Q}_{VV+AA}^{\Delta
S=2}$ will not mix with any other operator of the theory.

\subsection{A convenient lattice discretization of $C_{K'{\cal Q}K}$}
\label{sec:BKLATT}

In view of the previous remarks, we are led to consider the lattice
regularized gauge model~(\ref{FULACT}) with valence quarks $q_u \equiv
u$, $q_{d}\equiv d$, $q_{d'}\equiv d'$, $q_{s}\equiv s$, $q_{s'}\equiv
s'$ and $q_c = c$, each having an action of the form~(\ref{VALQACT})
(accompanied by the corresponding ghosts, $\phi_{u}$, $\phi_{d}$,
$\phi_{d'}$, $\phi_{s}$, $\phi_{s'}$, $\phi_{c}$) plus two sea quark
doublets, $\psi_\ell \equiv (u_{\rm sea},d_{\rm sea})$ and $\psi_h
\equiv (s_{\rm sea},c_{\rm sea})$.  The details of the lattice
formulation of this theory are as explained in general in
section~\ref{sec:VQ}. We will refer to this regularized theory as the
$(4s6v)_{tm}^{\rm L}$ model.

For reasons that will become soon clear we take the Wilson parameters
of the valence down and strange fields to be related in the following
way\footnote{Other equivalent choices, like $r_d=r_s=-r_{d'}=r_{s'}$
are possible. All the arguments we give in this section can be
immediately adapted to these other cases.}
\beq 
r_d=r_s=r_{d'}=-r_{s'} \, . \label{RFBK} 
\eeq
Obviously, in order to obtain from the $(4s6v)_{tm}^{\rm L}$ model
correlators that in the continuum limit are equal to those of
four-flavour euclidean QCD, we must require the renormalized
masses of the (sea and valence) quarks of the same flavour to take
identical values, i.e.\footnote{Our notation is such that
e.g.\ $m_\ell^{+} = m_{d_{\rm sea}}$ and $m_h^{-} = m_{s_{\rm
sea}}$.} 
\beq
\label{SDQMMATCHLAT} 
\begin{array}[b]{rclrcl}
\hat{m}_\ell^{-} &=& \hat{m}_u\, , \qquad &
\hat{m}_\ell^{+} &=& \hat{m}_d = \hat{m}_{d'} \, ,
\\[3pt]
\hat{m}_h^{-} &=& \hat{m}_s = \hat{m}_{s'} \, , \qquad&
\hat{m}_h^{+} &=& \hat{m}_c \, ,
\end{array}
\eeq 
and be equal to those of the
target theory, (qcd4). In terms of bare masses, while the conditions
$\hat{m}_d = \hat{m}_{d'}$ and $\hat{m}_s = \hat{m}_{s'}$ are
satisfied (owing to eqs.~(\ref{VALMQRE}) and~(\ref{RFBK}) and the fact 
that $Z_m(r_f)$ is even in $r_f$) by simply setting 
\beq 
\label{MFBK}
m_{d'} = m_d \, , \qquad m_{s'} = m_s \, , 
\eeq
the remaining four conditions in eq.~(\ref{SDQMMATCHLAT}) imply a
non-trivial one-to-one correspondence between $m_u$, $m_d$, $m_s$,
$m_c$ and the sea quark bare mass parameters $m_\ell$,
$\epsilon_\ell$, $m_h$, $\epsilon_h$, as dictated by
eqs.~(\ref{SEAMQRE})--(\ref{VALMQRE}).

We now prove that in the lattice model $(4s6v)_{tm}^{\rm L}$, with
the set of quark masses, $M$, satisfying the
conditions~(\ref{SDQMMATCHLAT}), the (bare) $B_K$ parameter can be
extracted from the correlator 
\beq 
a^6 \sum_{{\bf x},{\bf y}}
C_{K'{\cal Q}K}^{(4s6v)_{tm}^{\rm L}}(x,y) = a^6 \sum_{{\bf
x},{\bf y}} \langle (\bar{d}'\gamma_5 s')(x) \, 2 {\cal
Q}_{VV+AA}^{\Delta S=2}(0) (\bar{d}\gamma_5 s)(y)
\rangle\Big{|}_{(R,M)}\,, 
\label{COR2BKLAT} 
\eeq
with the operator ${\cal Q}_{VV+AA}^{\Delta S=2}$ given in
eq.~(\ref{OPBKCONT}) and the (m.r.) pseudo-scalar $(\bar{d}'\gamma_5
s')$ and $(\bar{d}\gamma_5 s)$ densities\footnote{Mixing of
$(\bar{d}'\gamma_5 s')$ with $(\bar{d}' s')$ and of $(\bar{d} \gamma_5
s )$ with $(\bar{d} s )$ is ruled out by the invariance of the lattice
model under ${\cal P}_5 \times (M \to - M)$, where ${\cal P}_5$ is
defined in eq.~(\ref{P5}). More precisely these are O($a$) effects
that appear with coefficients proportional to $a(m_s + m _d)$ or
$a(m_s - m_d)$.} playing the role of interpolating fields for the
$K^0$ and $\bar{K}^0$ states. The proof is divided in two parts. We
first show how $B_K$ is related to the large time behaviour of the
correlator~(\ref{COR2BKLAT}) and in the next section we prove that the
operator~(\ref{OPBKCONT}) does not mix with any other operator with
the same unbroken quantum numbers. A similar result was already
derived within the standard tm-LQCD approach of ref.~\cite{TMLQCD} and
exploited in ref.~\cite{GHPSV} in actual simulations. In the present
scheme one gets two extra bonuses: an automatically O($a$) improved
determination of $B_K$ (as it follows from the arguments we give at
the end of this section) and a positive definite fermion determinant
even in the mass non-degenerate case~\cite{FR2}.

On a lattice with very large temporal extension the leading contribution
to the spectral representation of the correlator~(\ref{COR2BKLAT})
in the $(4s6v)_{tm}^{\rm L}$ theory is given by
\beq 
\frac{e^{-M_{K'}|x_0|-M_{K}|y_0|} }{ 4M_{K'}M_K }
\langle\Omega|\bar{d}'\gamma_5
s'|\bar{K'}^{0}\rangle\langle\bar{K'}^0| 2 {\cal
Q}_{VV+AA}^{\Delta S=2} |K^0 \rangle \langle  K^0| \bar{d}\gamma_5
s | \Omega \rangle \biggr|_{(R,M)} \, , 
\eeq
where $K^0$ ($\bar{K'}^{0}$) is the neutral kaon (anti-kaon) state
created from the vacuum $\Omega$ by the operator $-\bar{d}\gamma_5 s$
($\bar{s'} \gamma_5 d'$) and $M_{K}$ ($M_{K'}$) the corresponding
meson mass. We remark that with the choice~(\ref{RFBK}) of the
(valence) Wilson parameters, the lattice kaon mass, $M_K$, and kaon
decay constant, $F_K$, defined in terms of the valence quark fields
$d$ and $s$ may differ by O($a^2$) effects from their counterparts,
$M_{K'}$ and $F_{K'}$, analogously defined in terms of the twin
valence quarks $d'$ and $s'$. Keeping track of this difference, we
write the relation which defines the bare $B_K$ parameter in the form
\beq
\langle \bar{K'}^0 | \, 2 {\cal Q}_{VV+AA}^{\Delta S=2} | K^0
\rangle = {16 \over 3} M_{K'} F_{K'} M_K F_K  B_K^{\rm bare} \, .
\label{BKDEFLAT} 
\eeq
The values of $M_K$, $F_K$, $M_{K'}$ and
$F_{K'}$ can be obtained in the usual way from the study of the
two-point correlation functions
\beq
a^3 \sum_{\bf x} \langle
(\bar{s}\gamma_5 d)(0) (\bar{d}\gamma_0\gamma_5 s)(x) \rangle \, ,
\qquad
a^3 \sum_{\bf x} \langle (\bar{s}'\gamma_5 d')(0)
(\bar{d}'\gamma_0 \gamma_5 s')(x) \rangle \, . 
\label{BK_AUXCORR}
\eeq
The renormalized $B_K$ parameter is finally given by 
\beq 
B_K(\mu)
= Z_{{\cal Q}_{VV+AA}^{\Delta S=2}}(a\mu) B_K^{\rm bare} \, ,
\label{BK_REN} 
\eeq
with $Z_{{\cal Q}_{VV+AA}^{\Delta S=2}}$ the
renormalization constant of ${\cal Q}_{VV+AA}^{\Delta S=2}$.

\pagebreak[3]

The correlation function~(\ref{COR2BKLAT}) can be O($a$) improved
via $WA$, according to the prescription~(\ref{WILAV}), and the
same holds for all masses and matrix elements that occur in its
spectral representation. However, this is not necessary. In fact,
thanks to the spurionic symmetry~(\ref{PSP}), $M_{K}$, $M_{K'}$,
$F_{K}$ and $F_{K'}$ (see eqs.~(\ref{ENWAS}) and~(\ref{MEWAS})),
as well as all the quantities that can be extracted from the
correlator~(\ref{COR2BKLAT}), are automatically free of O($a$)
cut-off effects. This property follows from the fact
that~(\ref{COR2BKLAT}) is the expectation value of a m.r.\
operator with the insertion of two other operators taken at
vanishing spatial three-momentum. Consequently, $B_K^{\rm bare}$
as defined above can be extracted from automatically O($a$)
improved data. The same is true for the renormalization constant of the
operator ${\cal Q}_{VV+AA}^{\Delta S=2}$ (see section~3.3 of
ref.~\cite{FR1} and section~\ref{sec:IRSS} of this paper).

\subsection{Proof of the absence of mixing in ${\cal Q}_{VV+AA}^{\Delta S=2}$}
\label{sec:BKNOMIX}

Absence of wrong chirality and parity mixing in the (renormalizable
expression of the) operator ${\cal Q}_{VV+AA}^{\Delta S=2}$ is proved
relying on the symmetries of the lattice model~(\ref{FULACT})
supplemented by the specifications given in
section~\ref{sec:BKLATT}. In particular the conditions~(\ref{RFBK})
and~(\ref{MFBK}) are assumed to be fulfilled.

It is almost obvious that the operator ${\cal Q}_{VV+AA}^{\Delta S=2}$
can not mix with operators of dimension less than six, because of its
flavour structure and the fact that the valence sector of the
action~(\ref{FULACT}) is diagonal in flavour. In particular, the
valence flavour quantum numbers $d$, $d'$, $s$ and $s'$ are all
conserved, as it formally follows from the invariance of the action
under the four $\U(1)$'s vector transformations, ${\cal I}_f$,
$f=d,d',s,s'$, of eq.~(\ref{FIXFLAROT}).  We conclude that ${\cal
Q}_{VV+AA}^{\Delta S=2}$ can only mix with four fermion operators with
$\Delta s = \Delta s' =1$ and $\Delta d = \Delta d' =-1$.

In order to rule out mixing with other six-dimensional operators
satisfying the above selection rules, one must use the following extra
symmetries enjoyed by the model action we are considering, i.e.\
\begin{enumerate}
\item ${\rm Ex}(d,d') \times (m_d \leftrightarrow m_{d'})$,
\item ${\rm Ex}_5(s,s') \times (m_s \leftrightarrow -m_{s'})$,
\item ${\cal C} \times [{\rm Ex}(d,s) \times (m_d \leftrightarrow m_{s})]
      \times  [{\rm Ex}_5(d',s') \times (m_{d'} \leftrightarrow -m_{s'})]$,
\item ${\cal P}_5 \times (M \rightarrow -M)$,
\end{enumerate}
where $\cal{C}$ and ${\cal P}_5$ are defined in
eqs.~(\ref{CHARGE}) and~(\ref{P5}) of appendix~\ref{sec:APPA}, respectively, and 
\beqn
{\rm Ex}(f_1,f_2) :&& q_{f_1} \leftrightarrow
q_{f_2} \quad\quad \bar{q}_{f_1}\leftrightarrow
\bar{q}_{f_2}\quad\quad\phi_{f_1}
\leftrightarrow \phi_{f_2} \, ,
\label{EX12}\\
{\rm Ex}_5(f_1,f_2) :&& 
\left \{\begin{array}{lll}
q_{f_1} \rightarrow \gamma_5 q_{f_2} & \quad
\bar{q}_{f_1} \rightarrow -\bar{q}_{f_2} \gamma_5
& \quad \phi_{f_2} \rightarrow \gamma_5 \phi_{f_1} 
\\
q_{f_2} \rightarrow \gamma_5 q_{f_1} & \quad
\bar{q}_{f_2} \rightarrow -\bar{q}_{f_1} \gamma_5
& \quad \phi_{f_1} \rightarrow \gamma_5 \phi_{f_2}
\end{array}\right .
\label{EX12_5} 
\eeqn 
The transformation ${\rm Ex}(f_1,f_2)$ simply
represents the exchange of valence flavours $f_1$ and $f_2$ 
with its action appropriately extended to ghost fields. The 
transformation ${\rm Ex}_5(f_1,f_2)$ is nothing but the product
\beq 
\label{EX12_5_bis} 
{\rm Ex}_5(f_1,f_2) = {\rm Ex}(f_1,f_2)
\times {\cal R}_{5f_1} \times {\cal R}_{5f_2}  \, . 
\eeq 
Notice that, if $r_{f_1} = r_{f_2}$, ${\rm Ex}(f_1,f_2)$ is a symmetry
of the action~(\ref{FULACT}) when combined with $(m_{f_1} \rightarrow
m_{f_2})$. If instead $r_{f_1} = - r_{f_2}$, the action is invariant
under ${\rm Ex}_5(f_1,f_2) \times (m_{f_1}\leftrightarrow -m_{f_2})$.

Symmetry~1 rules out the dimension six operators that,
besides obeying the selection rules $\Delta s=\Delta s'=-\Delta d
=-\Delta d'=1$, are not even under Ex($d$,$d'$). A basis of
operators with the desired unbroken quantum numbers is provided
by ${\cal Q}_{VV+AA}^{\Delta S=2}$ itself and the
operators~\cite{DELT6,CB} that, with obvious notations, we list below
\begin{equation}
\begin{array}[b]{lll}
{\cal Q}_{VV-AA}^{\Delta S=2} \, , \qquad&
  {\cal Q}_{SS\pm PP}^{\Delta S=2} \, , \qquad&
  {\cal Q}_{TT}^{\Delta S=2} \, ,
\\[3pt]
 {\cal Q}_{VA\pm AV}^{\Delta S=2} \, , \qquad&
  {\cal Q}_{SP\pm PS}^{\Delta S=2} \, , \qquad&
  {\cal Q}_{T\wt{T}}^{\Delta S=2} \, .
\end{array}
\label{OPLISTBK}
\end{equation}
For instance, we have ($\sigma_{\mu\nu}\equiv
i[\gamma_\mu,\gamma_\nu]/2$)
\begin{eqnarray}
{\cal Q}_{VA\pm AV}^{\Delta S=2} &=&
(\bar{s} \gamma_\mu d) (\bar{s}' \gamma_\mu\gamma_5 d')
\pm  (\bar{s} \gamma_\mu\gamma_5 d) (\bar{s}' \gamma_\mu d') +
\nonumber \\
&& + (\bar{s} \gamma_\mu d') (\bar{s}' \gamma_\mu\gamma_5 d)
\pm  (\bar{s} \gamma_\mu\gamma_5 d') (\bar{s}' \gamma_\mu d) \, ,
\label{OPBKNOT1} \\
{\cal Q}_{T\wt{T}}^{\Delta S=2} &=& \epsilon_{\mu\nu\lambda\rho}\Big{[}
(\bar{s} \sigma_{\mu\nu} d) (\bar{s}' \sigma_{\lambda\rho} d')
+ (\bar{s} \sigma_{\mu\nu} d') (\bar{s}' \sigma_{\lambda\rho}d)
\Big{]} \, .\qquad
\label{OPBKNOT2}
\end{eqnarray}
None of the (dimension six) operators in~(\ref{OPLISTBK}) can mix
with ${\cal Q}_{VV+AA}^{\Delta S=2}$ thanks to the other three (spurionic)
symmetries introduced above. In fact
\begin{itemize}
\item symmetry~2 rules out ${\cal Q}_{VV-AA}^{\Delta S=2}$ and ${\cal
      Q}_{SS-PP}^{\Delta S=2}$ (the latter are odd, while ${\cal
      Q}_{VV+AA}^{\Delta S=2}$ is even);
\item symmetry~3 rules out ${\cal Q}_{TT}^{\Delta S=2}$ and ${\cal
      Q}_{SS+PP}^{\Delta S=2}$ (the latter are odd, while ${\cal
      Q}_{VV+AA}^{\Delta S=2}$ is even);
\item symmetry~4 rules out ${\cal Q}_{VA\pm AV}^{\Delta S=2}$, ${\cal
      Q}_{SP\pm PS}^{\Delta S=2}$ and ${\cal Q}_{T\wt{T}}^{\Delta
      S=2}$ (the latter are odd, while ${\cal Q}_{VV+AA}^{\Delta S=2}$
      is even).
\end{itemize}
The symmetries we have employed to exclude mixing of ${\cal
Q}_{VV+AA}^{\Delta S=2}$ with the operators~(\ref{OPLISTBK}) can be
viewed as the analog of the usual (spurionic) CPS symmetry~\cite{CB}
translated in the present chirally twisted lattice framework. Indeed,
if valence quark (and ghost) fields are rotated to the basis where the
(subtracted) Wilson term appears in the standard (untwisted) form,
while the mass term is chirally twisted, the spurionic symmetry
obtained by combining the transformations~3 and~4 above turns out to
correspond to the CPS invariance of the standard Wilson theory.

We conclude this section by observing that the whole argument about
the absence of mixing in the operator entering the matrix element from
which $B_K$ is extracted could be repeated by duplicating only the
valence strange quark, and not the down quark. If one does so, one
also has to change the form of the operator in eq.~(\ref{OPBKCONT})
with the substitution
\beq 
{\cal Q}_{VV+AA}^{\Delta S=2} \to (\bar{s}\gamma_\mu d)
(\bar{s}'\gamma_\mu d)+(\bar{s} \gamma_\mu\gamma_5
d)(\bar{s}'\gamma_\mu\gamma_5 d) \label{REPT}
\eeq
and accordingly modify the expression of the interpolating field of
$K_0'$ from $\bar d'\gamma_5 s'$ to $\bar d \gamma_5 s'$. The reason
why this is possible lies in the observation that, given
eqs.~(\ref{RFBK}) and~(\ref{MFBK}), the flavour $d'$ is actually
indistinguishable from $d$.\footnote{We thank D. Be\'cirevi\'c for
drawing our attention on this point.} In section~\ref{sec:BKNOMIX} we
have presented the line of arguments leading to absence of mixing in
the apparently more complicated setting where also the down quark is
replicated, because it turns out that in this way the proof goes
through in a somewhat simpler fashion.

\section{The $\Delta I=1/2$ rule from $K\to\pi\pi$ amplitudes}
\label{sec:DIUM2}

A long standing puzzle in low energy particle physics, often referred to 
as ``octet enhancement'' or the $\Delta I = 1/2$ rule~\cite{GAS}, is
represented by the surprisingly large experimental value of the
ratio
\beq
{\rm R}(K\to\pi\pi) = {\Gamma(K\to \pi\pi)\vert_{\Delta
I=1/2} \over \Gamma(K\to \pi\pi)\vert_{\Delta I=3/2}}\, \sim \,
400\, .\label{DELI}
\eeq
Although the rate of the $\Delta I=3/2
\,\,K\to \pi\pi$ weak decays can be reasonably well computed
within our present understanding of field theory (based on O.P.E.
and renormalization group arguments), theoretical estimates of the
$\Delta I=1/2$ amplitude seem to give a much too small value 
compared to the experimental number~\cite{BB}.

\looseness=-1Due to the amount of seemingly well founded theoretical ideas involved
in this calculation, it is a crucial question to understand whether
this discrepancy is due to our ignorance of non-perturbative QCD
effects in the hadronic matrix elements of the composite quark
operators appearing in the effective weak hamiltonian, or whether this
problem should be rather taken as an indication of the existence of
some kind of new physics waiting to be uncovered in kaon
phenomenology. Lattice gauge theory represents the ideal framework in
which this question should find an answer. There are, however, severe
technical and conceptual difficulties in the process of establishing
the proper strategy to accomplish this task.

First of all, when working with euclidean metric, as one always does
in lattice simulations, the procedure necessary to extract the kaon
decay amplitudes of physical interest is complicated by IR subtleties
arising from final (two-pion) state interactions~\cite{MT}. To attack
this problem new interesting ideas have been recently put
forward~\cite{MT_WAYOUT,LMST}. Here we will not be concerned with such
delicate issues.

The main objective of this paper is rather to deal with the UV
difficulties related to the construction of the renormalized effective
weak hamiltonian operator on the lattice. We will show that tm-LQCD at
maximal twist offers a computational scheme in which the problem of
wrong chirality mixing, which afflicts the approach based on standard
Wilson fermions~\cite{BMMRT,DGMSTV},\footnote{See, however,
ref.~\cite{MRTTSS} for some way-outs if the Wilson theory is fully
O($a$) improved \`a la Symanzik~\cite{SYM,SW,HMPRS,REST}.} is
completely absent. An important step in this direction is represented
by the recent paper of ref.~\cite{PSV}, where the problem of mixing
has been brought to an amenable solution. Here we prove that the
calculation can be carried out in a fully O($a$) improved way without
the need of introducing the Sheikholeslami-Wohlert~\cite{SW} term in
the lattice action, or improving the lattice
operators~\cite{HMPRS,REST} representing the CP-conserving effective
weak hamiltonian, while still having a positive definite determinant
even for non-degenerate quarks.

\pagebreak[3]

\subsection{Kaon decay amplitudes}
\label{sec:KDA}

In the Standard Model the decay of the $K$ meson into pion states is 
described to leading order in the Fermi constant, $G_{\rm F}$, by the
CP-conserving, $\Delta S=1$ effective weak hamiltonian, which (in
the chiral limit) reads~\cite{GAS}
\beq
{\cal{H}}_{\rm{eff}}^{\Delta S=1} = V_{ud}V_{us}^* { G_{\rm F}
\over \sqrt{2}} \biggl[ C_{+}\left(\frac{\mu}{M_{\rm W}}\right)
\widehat{\cal O}^{+}(\mu) + C_{-}\left(\frac{\mu}{M_{\rm W}}\right)
\widehat{\cal O}^{-}(\mu) \biggr]\, , \label{H_eff_LL}
\eeq
where $V_{ud}V_{us}^*$ is the product
of the appropriate matrix elements of the CKM matrix.\footnote{As
usual, the top quark contribution, which is down by a factor
O($V_{td}V^\star_{ts}/V_{ud}V^\star_{us}\simeq 10^{-3}$), is
neglected.} The effective operator
${\cal{H}}_{\rm{eff}}^{\Delta S=1}$ is obtained after having
integrated out all degrees of freedom above some energy scale,
$\Lambda$, with $\Lambda$ larger than the charm threshold, but
still well below the W-boson mass, $M_{\rm W}$ Consistently, the
operators $\widehat{\cal O}^{\pm}$ in eq.~(\ref{H_eff_LL}) are
renormalized at the scale $\mu$ with $m_c\ll\mu\ll M_{\rm W}$, and
the Wilson coefficients $C_{\pm}(\mu/M_{\rm W})$ carry the
information about the physics between $\mu$ and $M_{\rm W}$. The
expression of the bare operators corresponding to $\widehat{\cal
O}^{\pm}$ is 
\beq 
{\cal O}^{\pm} = \half \Big[
(\bar{s}\gamma_\mu^L u) (\bar{u} \gamma_\mu^L d) \pm
(\bar{s}\gamma_\mu^L d) (\bar{u} \gamma_\mu^L u) \Big] - 
\half \Big[ u \leftrightarrow c \Big] \label{O_pm_LL} 
\eeq
with
$\gamma_\mu^L = \gamma_\mu (1-\gamma_5)$.

For the purpose of making contact with experimental data it is
enough to consider the decay of the neutral kaon, $K^0$, into
either $\pi^+ \pi^-$ or $\pi^0\pi^0$ states~\cite{CB,MMRT,HL,PSV}.
Owing to the parity invariance of the \emph{formal} continuum QCD,
the relevant amplitudes
\begin{eqnarray}
{\cal A}_{K^0 \to \pi^+\pi^-} &=&  V_{ud}V_{us}^*
{ G_{\rm F} \over \sqrt{2}} \sum_{j=+,-} C_j\left(\frac{\mu}{M_{\rm W}}\right)
\langle \pi^+\pi^- | \widehat{\cal O}^{\; j}_{VA+AV}(\mu) | K^0
\rangle \label{ME_pi+pi-} \, ,\qquad
\\
{\cal A}_{K^0 \to \pi^0\pi^0} &=&  V_{ud}V_{us}^*
{ G_{\rm F} \over \sqrt{2}} \sum_{j=+,-} C_j\left(\frac{\mu}{M_{\rm W}}\right)
\langle \pi^0\;\pi^0 |\, \widehat{\cal O}^{\; j}_{VA+AV}(\mu) |
K^0 \rangle \label{ME_pi0pi0} \, ,
\end{eqnarray}
can be written in terms of matrix elements of renormalized parity
odd operators,\linebreak $\widehat{\cal O}^{\pm}_{VA+AV}(\mu)$, with the
corresponding bare operators given by
\beqn
{\cal
O}^{\pm}_{VA+AV} &=& \half \Big[ (\bar{s}\gamma_\mu u) (\bar{u}
\gamma_\mu\gamma_5 d) \pm (\bar{s}\gamma_\mu d) (\bar{u}
\gamma_\mu\gamma_5 u) \Big]  -  \half \Big[ u \leftrightarrow c
\Big] +
\nonumber \\
&&+ \half \Big[ (\bar{s}\gamma_\mu\gamma_5 u) (\bar{u} \gamma_\mu
d) \pm (\bar{s}\gamma_\mu\gamma_5 d) (\bar{u} \gamma_\mu u) \Big]
 -  \half \Big[ u \leftrightarrow c \Big] \, .\qquad
\label{O_pm_VAAV} 
\eeqn

\subsection{Correlation functions and valence quarks}
\label{sec:CFVQ}

Information about the four matrix elements appearing in
eqs.~(\ref{ME_pi+pi-}) and~(\ref{ME_pi0pi0}) can be
extracted from suitable spatial Fourier transforms\footnote{To
get the kaon and the two-pion states at zero (total)
three-momentum one has to sum the
correlators~(\ref{K2PIC_contcorr})--(\ref{K2PIN_contcorr}) over
${\bf y}$ and $({\bf x} + {\bf z})/2$. In this way the two-pion
states can be labeled by the relative three-momentum of the two
pions.} of the correlators ($x_0>0$, $z_0>0$ and $y_0 <0$) 
\beqn 
C_{\pm , K^0\pi^+\pi^-}^{({\rm qcd}4)}(x,z,y) &=& \langle
\Phi_{\pi^+}(x) \Phi_{\pi^-}(z) {\cal O}^{\pm}_{VA+AV}(0)
\Phi^\dagger_{K^0}(y) \rangle_{\rm conn} \, ,
\label{K2PIC_contcorr} \\
C_{\pm , K^0\pi^0\pi^0}^{({\rm qcd}4)}(x,z,y) &=& \langle
\Phi_{\pi^0}(x) \Phi_{\pi^0}(z) {\cal O}^{\pm}_{VA+AV}(0)
\Phi^\dagger_{K^0}(y) \rangle_{\rm conn} \, ,
\label{K2PIN_contcorr} 
\eeqn 
where with the notation $\langle O \rangle_{\rm conn}$ we denote the
connected part of the vacuum expectation value of the multi-local
operator $O$. As before, the superscript $({\rm qcd}4)$ is to remind
us that correlators are understood to be evaluated in the formal
continuum QCD with four quark flavours ($u$, $d$, $s$ and
$c$). $\Phi_{K^0}$, $\Phi_{\pi^+}$, $\Phi_{\pi^-}$ and $\Phi_{\pi^0}$
are the interpolating operators for kaon and pion states,
respectively.

Our approach here is very similar in spirit to that we have
proposed in section~\ref{sec:BK} for the evaluation of $B_K$. In
order to explain the idea it is convenient to first consider at a
formal level a gauge model with, as usual, four sea quark flavours
($u_{\rm sea}$, $d_{\rm sea}$, $s_{\rm sea}$ and $c_{\rm sea}$),
plus several (actually ten) valence quark species. For reasons
that will become clear in the following, each of the up and charm
valence quark species will have to be replicated four times. The
pattern of valence quarks we wish to consider is then [$u$, $u'$,
$u''$, $u'''$, $d$, $s$, $c$, $c'$, $c''$, $c'''$]. Each quark
species is introduced together with the ghost necessary to cancel
the loop contribution coming from the corresponding fermionic
functional integration. We will denote the correlators of this
model by the superscript $(4s10v)$. We now consider the
(connected) correlators 
\beqn 
C_{\pm ,K^0\pi^+\pi^-}^{(4s10v)}(x,z,y) &=& \langle \Phi_{\pi^+}(x)
\Phi_{\pi^-}(z) {\cal Q}^{\pm}_{VA+AV}(0) \Phi^\dagger_{K^0}(y)
\rangle_{\rm conn}\, ,
\label{K2PIC_modcorr} \\
C_{\pm , K^0\pi^0\pi^0}^{(4s10v)}(x,z,y) &=& \langle \Phi_{\pi^0}(x)
\Phi_{\pi^0}(z) {\cal Q}^{\pm}_{VA+AV}(0) \Phi^\dagger_{K^0}(y)
\rangle_{\rm conn} \, ,\label{K2PIN_modcorr} 
\eeqn
where the meson
interpolating operators $\Phi_{K^0}$, $\Phi_{\pi^+}$,
$\Phi_{\pi^-}$ and $\Phi_{\pi^0}$ are made out of the valence
quarks $u$, $d$ and $s$ (but not $u'$, $u''$ or $u'''$). This
fact will be crucial for the validity of the
relations~(\ref{CpmK2PIC_EQUAL})--(\ref{CpmK2PIN_EQUAL}) below.
The key point in eqs.~(\ref{K2PIC_modcorr})--(\ref{K2PIN_modcorr})
is the introduction of the auxiliary operators ${\cal
Q}^{\pm}_{VA+AV}$, defined by the formula 
\beq 
{\cal
Q}^{\pm}_{VA+AV} = {\cal Q}^{\pm \, [0]}_{VA+AV}
 + {\cal Q}^{\pm \, [1]}_{VA+AV}
 - {1 \over 2} {\cal Q}^{\pm \, [2]}_{VA+AV}
 - {1 \over 2} {\cal Q}^{\pm \, [3]}_{VA+AV} \, .
\label{Q_01_23_COMB} 
\eeq
where the operators 
${\cal Q}^{\pm \,[k]}_{VA+AV}$, $k=0,1,2,3$ 
are all of the form~(\ref{O_pm_VAAV}), but differ from each other
by the type of $u$ and $c$ valence quark species they contain.
Explicitly we have
\beqn
{\cal Q}^{\pm
\,[k]}_{VA+AV} &=& \half \Big[ (\bar{s}\gamma_\mu u^{[k]} ) (\bar{u}^{[k]}
\gamma_\mu\gamma_5 d ) \pm (\bar{s}\gamma_\mu d )
(\bar{u}^{[k]} \gamma_\mu\gamma_5 u^{[k]} ) \Big]  -  \half \Big[ u^{[k]}
\leftrightarrow c^{[k]} \Big] + 
\nonumber \\&&
+\half \Big[ (\bar{s}\gamma_\mu\gamma_5 u^{[k]} ) (\bar{u}^{[k]} \gamma_\mu
d ) \pm (\bar{s}\gamma_\mu\gamma_5 d ) (\bar{u}^{[k]}
\gamma_\mu u^{[k]} ) \Big]  -  \half \Big[ u^{[k]} \leftrightarrow c^{[k]}
\Big] \qquad
\label{Q_pm_VAAV_l} 
\eeqn
with
\beq
\begin{array}[b]{rclrclrclrcl} 
u^{[0]} &\equiv& u \, ,
\qquad& u^{[1]} &\equiv& u' \, , \qquad& u^{[2]} &\equiv& u'' \,
, \qquad& u^{[3]} &\equiv& u''' \, ,
\\[3pt]
c^{[0]} &\equiv& c \, , \qquad& c^{[1]} &\equiv& c' \, ,
\qquad& c^{[2]} &\equiv& c'' \, , \qquad& c^{[3]} &\equiv& c'''
\, . \label{VALREPLNOT}
\end{array}
\eeq
As argued in the previous section, if the two models $(4s10v)$ and
$({\rm qcd}4)$ are regularized in the same way (say, using GW
fermions), a straightforward application of the Wick theorem leads to the
equalities
\beqn
C_{\pm ,K^0\pi^+\pi^-}^{(4s10v)}(x,z,y) & = & C_{\pm , K^0\pi^+\pi^-}^{({\rm
qcd}4)}(x,z,y)\, , \label{CpmK2PIC_EQUAL} \\
C_{\pm , K^0\pi^0\pi^0}^{(4s10v)}(x,z,y) & =
& C_{\pm , K^0\pi^0\pi^0}^{({\rm qcd}4)}(x,z,y)\, ,
\label{CpmK2PIN_EQUAL} 
\eeqn valid at any finite value of the
lattice spacing, provided bare (and thus also renormalized) masses
of quarks having the same flavour are given the same value in the
two theories.
Eqs.~(\ref{CpmK2PIC_EQUAL})--(\ref{CpmK2PIN_EQUAL}) in particular
imply that the renormalization constants of the operator ${\cal
Q}^{\pm}_{VA+AV}$ in $(4s10v)$ can be taken equal to that of the
operator ${\cal O}^{\pm}_{VA+AV}$ in $({\rm qcd}4)$. An analogous
result is obtained also for the renormalization constants of the
meson interpolating operators
in the two models. Under these renormalization conditions, the
equalities~(\ref{CpmK2PIC_EQUAL})--(\ref{CpmK2PIN_EQUAL}) hold for
the corresponding renormalized correlators and carry over to the 
continuum limit.

We conclude this section by offering an intuitive explanation
of why we have introduced four copies of the $u$ and $c$ valence
quarks to get all the mixings canceled. The reason is the
following. For the purpose of killing the operator mixing coming
from ``penguin-like diagrams'' it would have been enough to add to the
primary $u$ and $c$ quarks only one extra replica of each of them, 
$u'$ and $c'$, with $r_{u'}=-r_{u}$ and $r_{c'}=-r_{c}$ (see also 
section~\ref{sec:lowdimmix} below). The whole pattern is then duplicated 
in order to have on each gauge background the penguin-like and the 
non-penguin-like contractions with the correct relative multiplicity 
factor.

\subsection{A convenient lattice discretization of the
$C_{\pm,K\pi\pi}^{(4s10v)}$ correlators}
\label{sec:CLDC}

In view of the previous considerations 
a convenient regularization of the $(4s10v)$ model is obtained by
taking the action~(\ref{FULACT}) with valence quarks $q_{u}\equiv u$,
$q_{u'}\equiv u'$, $q_{u''}\equiv u''$, $q_{u'''}\equiv u'''$,
$q_{d}\equiv d$, $q_{s}\equiv s$, $q_{c}\equiv c$,
$q_{c'}\equiv c'$, $q_{c''}\equiv c''$ and $q_{c'''}\equiv c'''$,
each having an action of the form~(\ref{VALQACT}) (accompanied by
the corresponding ghosts, $\phi_{u}$, $\phi_{u'}$, \dots, $\phi_{d}$, 
$\phi_{s}$, $\phi_{c}$, $\phi_{c'}$, \dots), plus, as always, the two 
usual sea quark doublets,
$\psi_\ell \equiv (u_{\rm sea},d_{\rm sea})$ and
$\psi_h \equiv (s_{\rm sea},c_{\rm sea})$. For reasons which
have to do with the need of making as
simple as possible the renormalization pattern of the operators
${\cal Q}^{\pm}_{VA+AV}$, we take the Wilson parameters of the
valence fields related as follows
\beqn
& r_d=r_s \, ,  
\nonumber \\
& r_u = -r_{u'} = r_{u''} =  -r_{u'''} = r_c = -r_{c'} = r_{c''} =
-r_{c'''} \, . \label{RF_Kto2PI} 
\eeqn
For definiteness, we also fix the relation between the two sets
of Wilson parameters in eq.~(\ref{RF_Kto2PI}) by setting
\beq r_u = r_d \, . \label{RF_pions_NEW} 
\eeq
The choice $r_u=-r_d$ is also possible. We comment in
appendix~\ref{sec:APPB} on the implications of this (isospin breaking)
choice.

The euclidean lattice theory we have just described (see also
section~\ref{sec:VQ}) will be referred to in the following as the
$(4s10v)^{\rm L}_{tm}$ model.

In order for the correlators of the $(4s10v)^{\rm L}_{tm}$ model (with
no insertion of sea or ghost fields) to be equal to those of (qcd4),
we must require the renormalized masses of sea and valence fields to
be matched as follows
\beqn
\hat{m}_\ell^{-} &=& \hat{m}_{u} = \hat{m}_{u'} =
\hat{m}_{u''} = \hat{m}_{u'''} \, , \qquad
\hat{m}_\ell^{+} = \hat{m}_d \, ,
\label{MUDCS_MATCH_LAT1} \\
\hat{m}_h^{+} &=& \hat{m}_{c} = \hat{m}_{c'} = \hat{m}_{c''} =
\hat{m}_{c'''} \, , \qquad \hat{m}_h^{-} = \hat{m}_s \, .
\label{MUDCS_MATCH_LAT2} 
\eeqn  
Owing to eqs.~(\ref{SEAMQRE})--(\ref{VALMQRE})
and the equality in absolute value of all the valence quark $r$ 
parameters (eq.~(\ref{RF_Kto2PI})), 
the formulae~(\ref{MUDCS_MATCH_LAT1})--(\ref{MUDCS_MATCH_LAT2}) 
fix the non-trivial relations between the bare mass parameters of 
sea and valence quarks and imply the bare masses of valence quarks 
of the same flavour to be equal, i.e.\  
\beq
\label{MUC_BARE_EQ} m_{u'''}=m_{u''}=m_{u'}=m_u \, , \qquad
m_{c'''}=m_{c''}=m_{c'}=m_c \, . 
\eeq 
As for the interpolating meson fields, one can use the standard 
pseudo-scalar quark bilinears (see, however, appendix~\ref{sec:APPB})
\beqn
&\Phi^\dagger_{K^0}= -\bar{d} \gamma_5 s \, ,
\quad \Phi_{\pi^+} = -\bar{d} \gamma_5 u \, ,
\quad \Phi_{\pi^-} =\Phi^\dagger_{\pi^+} =\bar{u} \gamma_5 d \, ,
\label{ANNIH_KIFIELD}\\
&\displaystyle\Phi_{\pi^0} = -\Phi^\dagger_{\pi^0} = {1 \over
\sqrt{2}} \Big[ \bar{u} \gamma_5 u - \bar{d}
\gamma_5 d \Big] \, .\label{ANNIH_PFIELD}
\eeqn

The fundamental reason for considering the $(4s10v)^{\rm L}_{tm}$
model with its somewhat weird valence quark content is that under the
conditions~(\ref{RF_Kto2PI}) and~(\ref{MUC_BARE_EQ}) the $\Delta S=1$
operators ${\cal Q}^{\pm}_{VA+AV}$ of eq.~(\ref{Q_01_23_COMB}) mix
neither between themselves nor with any operator of wrong chirality or
parity. This result will be proved in section~\ref{sec:NOMIX_O_VAAV}
below by exploiting the symmetries of the $(4s10v)^{\rm L}_{tm}$
model. As for the mixing with the famous three-dimensional operator
$\bar{s}\gamma_5 d$, we will show that, just like in the formal
chirally invariant continuum theory, it enters with a factor $(m_c^2 -
m_u^2)(m_s-m_d)$, leaving no room for any power divergent behaviour.
Similarly, despite the breaking of parity at non-zero lattice spacing,
no mixing with the scalar density operator $\bar{s} d$ occurs.

It should also be recalled that in evaluating zero momentum transfer
matrix elements of $\widehat{\cal Q}^{\pm}_{VA+AV}$ (like
$K\to\pi\pi$), the operator $\bar{s} \gamma_5 d$ does not play any
role, because PCAC allows to write $\bar{s}\gamma_5
d=\partial_\mu(\bar{s}\gamma_\mu\gamma_5 d)/(m_s + m_d)$ from which we
see that, up to contact terms, $\bar{s}\gamma_5 d$ can always be
replaced by a total divergence.

At this point we can conclude that, after subtraction of the
logarithmic divergent mixing of ${\cal Q}^{\pm}_{VA+AV}$ with $(m_c^2
- m_u^2)(m_s-m_d) \bar{s}\gamma_5 d$ (whenever needed), and
multiplication by the appropriate renormalization constants, the
correlators~(\ref{K2PIC_modcorr}) and~(\ref{K2PIN_modcorr}), evaluated
in the $(4s10v)^{\rm L}_{tm}$ model, admit a well defined continuum
limit. The limiting quantities, under the
conditions~(\ref{MUDCS_MATCH_LAT1})--(\ref{MUDCS_MATCH_LAT2}),
coincide with the (continuum limit of the)
correlators~(\ref{K2PIC_contcorr})--(\ref{K2PIN_contcorr}) in the
$({\rm qcd}4)$ model, thanks to the
equalities~(\ref{CpmK2PIC_EQUAL})--(\ref{CpmK2PIN_EQUAL}).

\subsection{Some observations}
\label{sec:SCR1}

A number of observations are in order at this point. First of all, we
stress that, on the basis of the general arguments given in
section~\ref{sec:VQ}, the
correlators~(\ref{K2PIC_modcorr})--(\ref{K2PIN_modcorr}) can be O($a$)
improved via $WA$'s (see eq.~(\ref{WILAV})). Actually, owing to the
symmetry under ${\cal P} \times (R \rightarrow -R)$ enjoyed by the
$(4s10v)^{\rm L}_{tm}$ model, the necessary $WA$'s of the matrix
elements of interest can be constructed in terms of data obtained from
a single simulation, as explained in section~\ref{sec:IRSS}, by
averaging --- whenever necessary --- matrix elements with opposite
value of the relative three-momentum of the two pions.

Secondly, for completeness we recall that, in order to actually
extract from simulation data the amplitudes ${\cal
A}_{K^0\to\pi^+\pi^-}$ and ${\cal A}_{K^0\to\pi^0\pi^0}$, it is also
necessary to evaluate some extra two- and four-point correlation
functions, such as
\beqn
&
\langle \Phi_{K^0}(x)
\Phi^\dagger_{K^0}(x') \rangle\, , \qquad \langle
\Phi_{\pi^+}(x) \Phi^\dagger_{\pi^+}(x') \rangle \, , \qquad
\langle \Phi_{\pi^0}(x) \Phi_{\pi^0}(x')
\rangle_{\rm conn}  \, ,
\nn  \\
&
\langle \Phi_{\pi^+}(x)
\Phi^{\dagger}_{\pi^+}(z) \Phi_{\pi^+}(x')
\Phi^{\dagger}_{\pi^+}(z') \rangle_{\rm conn} \, ,\qquad
\langle \Phi_{\pi^0}(x) \Phi_{\pi^0}(z) \Phi_{\pi^0}(x')
\Phi_{\pi^0}(z') \rangle_{\rm conn} \, ,
\label{KPIPI_AUXCORR2}
\eeqn
where in the last two four-point correlators one should take $x_0, z_0
>0$ and $x'_0, z'_0 <0$.  Suitable spatial Fourier transforms of these
correlators provide the necessary extra information on the energies of
the kaon, single pion and two-pion states, as well as on the
$\pi\pi$-phase shift~\cite{MT_WAYOUT,LMST}, which will in principle
enable us to extract the physical decay amplitudes of interest.
Finally, when not automatic, the correlators~(\ref{KPIPI_AUXCORR2})
can be O($a$) improved by taking appropriate $WA$'s.

Once the bare amplitudes of interest have been evaluated with no
O($a$) cut-off effects, a non-perturbative lattice estimate of the
renormalization constants, $Z_{VA+AV}^{\pm}(a\mu)$, is required.  The
latter can be obtained following the strategy outlined in section~3.3
of ref.~\cite{FR1}. One could e.g.\ consider a correlator where ${\cal
Q}^{\pm}_{VA+AV}$ is inserted at zero three-momentum with some other
local m.r.\ operator having appropriate quantum numbers so as not to
get an identically vanishing two-point function (see also
section~\ref{sec:IRSS}).

In the same spirit of the observation made at the end of
section~\ref{sec:BKNOMIX}, we note that one can eliminate the
flavour species $u'''$ and $c'''$ from the game (as they are
indistinguishable from $u'$ and $c'$, respectively) and change the
form of the operator in eq.~(\ref{Q_01_23_COMB}) with the
substitution 
\beq
{\cal Q}^{\pm}_{VA+AV} \to {\cal Q}^{\pm \,
[0]}_{VA+AV}+{1 \over 2} {\cal Q}^{\pm \, [1]}_{VA+AV} - {1 \over
2} {\cal Q}^{\pm \, [2]}_{VA+AV}\, .\label{REPTT}
\eeq 
Again, in this framework the proof of absence of mixing becomes
somewhat more complicated than the one we give below.

\subsection{The renormalization properties of ${\cal Q}^{\pm}_{VA+AV}$}
\label{sec:NOMIX_O_VAAV}

In this section we want to prove that the operators ${\cal
Q}^{\pm}_{VA+AV}$ (eq.~(\ref{Q_01_23_COMB})) do not mix with any
operator of lower or equal dimension, having either wrong chirality or
opposite parity, precisely as in the formal continuum theory or within
chirally invariant lattice regularizations (like those provided by
fermions obeying the GW condition). This in practice means that no
mixing with operators of the same or lower dimension is relevant for
the computation of the $K\to\pi\pi$ decay amplitudes. In the arguments
of this section a crucial role is played by the
conditions~(\ref{RF_Kto2PI}) and~(\ref{MUC_BARE_EQ}), which are always
assumed as part of the specification of the lattice $(4s10v)^{\rm
L}_{tm}$ model we consider.  Thanks to eq.~(\ref{RF_Kto2PI}) and the
condition $|r_u|=|r_d|$, the valence flavour symmetries of the
action~(\ref{FULACT}) rule out the mixing of ${\cal Q }^{+}_{VA+AV}$
and ${\cal Q }^{-}_{VA+AV}$ between themselves (see
appendix~\ref{sec:APPC} for details), as in the case of untwisted
Wilson fermions.  We can hence deal ``in parallel'' with ${\cal Q
}^{+}_{VA+AV}$ and ${\cal Q }^{-}_{VA+AV}$.

\subsubsection{Operators of dimension lower than six}
\label{sec:lowdimmix}

Gauge, $H(4)$ and ${\cal I}_f$ (with $f=s,d$, see eq.~(\ref{FIXFLAROT}))
invariances of the $(4s10v)^{\rm L}_{tm}$ lattice model
\pagebreak[3]
restrict the possible operators of dimension less than six which
${\cal Q}^{\pm}_{VA+AV}$ can mix with to only the following ones
\beq
\bar{s} \gamma_5 d\, , \qquad \bar{s} d \, , \qquad
\bar{s} \gamma_5 \sigma \!\cdot\! F d\, , \qquad \bar{s} \sigma
\!\cdot\! F d \, , \label{MIX35_VAAV_BAS} 
\eeq
where $\sigma
\!\cdot\! F = \sigma_{\mu\nu} F_{\mu\nu}[U]$ and $F_{\mu\nu}[U]$
stands for any lattice discretization of the gluon field strength.

We wish to show now that, thanks to the (spurionic) symmetries enjoyed
by the $(4s10v)^{\rm L}_{tm}$ lattice model, the above operators can
only come into play multiplied by at least three powers of quark
masses, according to the pattern given below
\begin{eqnarray}
(m_c - m_u) (m_c + m_u) (m_s - m_d) \hphantom{(m_s + m_d)}
&\qquad& \bar{s} \gamma_5 d \, ,
\label{MIX3+3} \\
(m_c - m_u) (m_c + m_u) (m_s - m_d) (m_s + m_d)
&& \bar{s} d \, ,
\label{MIX3+4} \\
(m_c - m_u) (m_c + m_u) (m_s - m_d) \hphantom{(m_s + m_d)}
&& \bar{s} \gamma_5 \sigma \!\cdot\! F d \, ,
\label{MIX5+3} \\
(m_c - m_u) (m_c + m_u) (m_s - m_d) (m_s + m_d) 
&& \bar{s} \sigma \!\cdot\! F d \, . 
\label{MIX5+4} 
\end{eqnarray}
As a result, for obvious dimensional reasons none of the above
operators can appear multiplied by power divergent mixing
coefficients. Only the operator~(\ref{MIX3+3}) will enter with a
logarithmically divergent coefficient. However, as we have already
discussed in section~\ref{sec:CLDC}, it gives vanishing contributions
to the physical decay amplitudes. One might add that the mass factor
in front of it vanishes anyway in the chiral limit or in the limit of
exact $\SU_V(3)$.

Since the operators $\bar{s} \gamma_5 d$ and
$\bar{s} \gamma_5 \sigma \!\cdot\! F d$ (as well as
$\bar{s} d$ and $\bar{s} \sigma \!\cdot\! F d$) have identical transformation
properties under all the symmetries we will be using in the argument that
follows, it is sufficient to restrict our attention to the
operators~(\ref{MIX3+3}) and~(\ref{MIX3+4}). Consistently with this
observation we notice that the overall quark mass factor
in front of the operators~(\ref{MIX5+3}) and~(\ref{MIX5+4})
is the same as that appearing in front of the operators~(\ref{MIX3+3})
and~(\ref{MIX3+4}), respectively.

We give a proof of the structure of the mass dependence displayed in
eqs.~(\ref{MIX3+3}) and~(\ref{MIX3+4}) in four steps.
\begin{enumerate}
\item The presence of the factor $(m_c-m_u)$ follows from the
spurionic invariance of the lattice action~(\ref{FULACT}) of the
$(4s10v)^{\rm L}_{tm}$ model under the transformation
\beq
{\rm Ex}(u,c) \times {\rm Ex}(u',c')
\times {\rm Ex}(u'',c'') \times {\rm Ex}(u''',c''') \times (m_u
\leftrightarrow m_c) \, , \label{EXUC} 
\eeq 
where ${\rm Ex}(f_1,f_2)$ is defined in eq.~(\ref{EX12}). Indeed,
under~(\ref{EXUC}) the operators ${\cal Q}^{\pm}_{VA+AV}$ are odd,
while $\bar{s} \gamma_5 d$ and $\bar{s} d$ are both even, as they are
left untouched by the transformation.

\item The presence of the factor $(m_s-m_d)$, 
is a consequence of the spurionic invariance of the
action under the transformation
\beq
{\rm Ex}(d,s) \times {\cal C}
\times (m_d \leftrightarrow m_s) \, , \label{EXDS*CC} 
\eeq
where ${\cal C}$ is the charge conjugation operation
(eq.~(\ref{CHARGE})). Under the transformation~(\ref{EXDS*CC}) the
operators ${\cal Q}^{\pm}_{VA+AV}$ are odd, while $\bar{s} \gamma_5 d$
and $\bar{s} d$ are even.

\item At this stage we have proven that both the operators
$\bar{s}\gamma_5 d$ and $\bar{s} d$ must appear multiplied by a factor
of the form $(m_c-m_u)(m_s-m_d)$.  Based on this result we now show
that a further factor $(m_c+m_u)$ must be present owing to the
spurionic invariance of the action under the transformation
\beqn
& {\rm Ex}_5(u,u') \times {\rm Ex}_5(u'',u''') \times (m_u
\rightarrow -m_u) \times
\nonumber \\
& \times \; {\rm Ex}_5(c,c') \times {\rm Ex}_5(c'',c''') \times
(m_c \rightarrow -m_c) \, , \label{EX01*EX23_5} 
\eeqn
where ${\rm Ex}_5(f_1,f_2)$ is defined in eq.~(\ref{EX12_5}). In fact,
under the transformation~(\ref{EX01*EX23_5}) the operators ${\cal
Q}^{\pm}_{VA+AV}$ are even, while the combinations $(m_c-m_u)\bar{s}
\gamma_5 d$ and $(m_c-m_u)\bar{s} d$ are odd.  Consequently, the mass
factor in front of both operators will have to have the form
$(m_c^2-m_u^2)(m_s-m_d)$. Thus the GIM suppression factor turns out to
be quadratic just like in the continuum.

We remark that each of the operators ${\cal Q}^{\pm\;[k]}_{VA+AV}$,
$k=0,1,2,3$, that make up ${\cal Q}^{\pm}_{VA+AV}$, actually mixes
with $(m_c-m_u)(m_s-m_d)\bar{s} \gamma_5 d$ and
$(m_c-m_u)(m_s-m_d)\bar{s} d$, with linearly divergent
coefficients. Owing to the spurionic invariance ${\cal
R}_{5u^{[k]}}^{\rm sp}\times{\cal R}_{5c^{[k]}}^{\rm sp}$
(eq.~(\ref{R5SPVAL})), such coefficients are, however, odd in
$r_{u^{[k]}} = r_{c^{[k]}}$. As a consequence of this property and the
peculiar valence flavour structure of the regularization we have
introduced (see eq.~(\ref{RF_Kto2PI})), all the mixings with lower
dimensional operators cancel against each other in the
combination~(\ref{Q_01_23_COMB}), which is symmetric under $(u,c)
\leftrightarrow (u',c')$ and $(u'',c'') \leftrightarrow (u''',c''')$.

\item On the basis of the further spurionic invariance
of the lattice action under the transformation 
\beq 
{\cal P}_5 \times (M \rightarrow -M) \, , \label{P5*Mrev} 
\eeq
where ${\cal P}_5$ is defined in eq.~(\ref{P5}), one concludes that
the mixing of ${\cal Q}^{\pm}_{VA+AV}$ with $\bar{s} d$ must be
suppressed by an extra quark mass factor, besides $(m_c^2-m_u^2)$ and
$(m_s-m_d)$. This is so because the operators ${\cal Q}^{\pm}_{VA+AV}$
are odd under the transformation~(\ref{P5*Mrev}), precisely as
$(m_c^2-m_u^2)(m_s-m_d)\bar{s}\gamma_5 d$, while the combination
$(m_c^2-m_u^2)(m_s-m_d)\bar{s} d$ is even. In front of the latter a
further mass factor must hence be present, which for consistency with
the symmetries~(\ref{EXUC}),~(\ref{EXDS*CC}) and~(\ref{EX01*EX23_5})
can only be of the form $(m_s+m_d)$. We get in this way the expression
anticipated in eq.~(\ref{MIX3+4}), which we notice is a mere O($a$)
cut-off effect. Its contribution actually cancels in Wilson-averaged
correlators and derived matrix elements.
\end{enumerate}

We have thus proved that the operators ${\cal Q}^{\pm}_{VA+AV}$ do
not mix with operators of dimension lower than six.

\subsubsection{Operators of dimension six}
\label{sec:nodim6mix}

The operators ${\cal Q}^{\pm}_{VA+AV}$ (eq.~(\ref{Q_01_23_COMB})) are
linear combinations of the operators ${\cal Q}^{\pm \, [k]}_{VA+AV}$,
defined in eq.~(\ref{Q_pm_VAAV_l}). The latter --- according to the
notation introduced in~(\ref{VALREPLNOT}) --- contain, besides the
valence $d$ and $s$ quarks, the flavour species $u$ and $c$ when
$k=0$, $u'$ and $c'$ when $k=1$, $u''$ and $c''$ when $k=2$ and
finally $u'''$ and $c'''$ when $k=3$.

The somewhat involved proof that the operators ${\cal Q}^{\pm \,
[k]}_{VA+AV}$, $k\in\{0,1,2,3\}$ and, hence, ${\cal Q}^{\pm}_{VA+AV}$
(eq.~(\ref{Q_01_23_COMB})) mix neither between themselves nor with any
other operator of dimension six is given in
appendix~\ref{sec:APPC}.\footnote{As we have already pointed out, the
operators ${\cal Q}^{\pm\,[k]}_{VA+AV}$ separately do mix with
operators of lower dimensions.  But in the relevant combination of
eq.~(\ref{Q_01_23_COMB}) the latter cancel out and arrange themselves
so as to give rise precisely to the pattern of eqs.~(\ref{MIX3+3})
to~(\ref{MIX5+4}), as was shown in section~\ref{sec:lowdimmix}.}

Incidentally we remark that the overall renormalization constant of
each ${\cal Q}^{\pm\, [k]}_{VA+AV}$ does not depend on the sign of
$r_{u^{[k]}} = r_{c^{[k]}}$ (see eq.~(\ref{RF_Kto2PI})), owing to the
general argument given below eq.~(\ref{RCONT}).  Since all the valence
Wilson parameters are taken to be equal in absolute value, it follows
that the renormalization constant of ${\cal Q}^{\pm \, [k]}_{VA+AV}$
is actually independent of $k$, i.e.\ is the same for all the terms
that make up ${\cal Q}^{\pm}_{VA+AV}$.

\vspace{-1pt minus 4pt}
\section{The $\Delta I=1/2$ rule from $K\to\pi$ amplitudes}
\label{sec:DIUM1}
\vspace{-1pt minus 2pt}

In the chiral limit $K\to\pi\pi$ amplitudes can be related to $K\to
\pi$ and $K\to vacuum$ amplitudes by the soft pion theorems
(SPT's)~\cite{LB} of Current Algebra~\cite{CA}.  The interest of these
formulae lies in the fact that they allow to evaluate in the chiral
limit $K\to\pi\pi$ amplitudes from the knowledge of $K\to\pi$ matrix
elements which are simpler to compute numerically because the final
(and the initial) state is a one-particle state and the no-go theorem
of ref.~\cite{MT} does not apply. Naturally, as these relations are
among amplitudes evaluated in the limit of vanishing quark masses, a
non-trivial extrapolation to the physical mass point will have to be
performed.

Starting again from eq.~(\ref{H_eff_LL}), it can be
proved~\cite{CB,MMRT,HL,PSV} that in QCD with four flavours it is
enough to compute (for a range of quark masses and momenta of the
kaon and pion states) the matrix elements 
\beq 
\langle\pi^+(p)|\widehat{\cal O}^{\pm}_{VV+AA}
|K^+(q)\rangle\, ,\qquad\langle\pi^0(p)|\widehat{\cal
O}^{\pm}_{VV+AA} |K^0(q)\rangle\, ,
\label{melkp} 
\eeq
where the
bare expression of the parity even operators $\widehat{\cal
O}^{\pm}_{VV+AA}$ reads
\beqn
{\cal O}^{\pm}_{VV+AA} &=& \half
\Big[ (\bar{s}\gamma_\mu u) (\bar{u} \gamma_\mu d) \pm
(\bar{s}\gamma_\mu d) (\bar{u} \gamma_\mu u) \Big]  -  \half
\Big[u
\leftrightarrow c \Big]  + 
\nonumber \\[-1pt minus 1pt]&& 
+ \half \Big[ (\bar{s}\gamma_\mu\gamma_5 u) (\bar{u}
\gamma_\mu\gamma_5 d) \pm (\bar{s}\gamma_\mu\gamma_5 d) (\bar{u}
\gamma_\mu\gamma_5 u) \Big] - \half \Big[ u \leftrightarrow c \Big] \,
.\qquad
\label{O_pm_VVAA}
\eeqn

Following a strategy analogous to the one developed in the
previous section, we want to prove that the matrix
elements~(\ref{melkp}) can be extracted (actually with no O($a$)
cut-off effects) from correlation functions defined in an auxiliary
lattice regularized theory where the relevant four fermion
operators do not mix with any other operator of wrong chirality or
opposite parity.

\vspace{-1pt minus 3pt}
\subsection{Correlation functions and valence quarks}
\label{sec:CFVQKP}
\vspace{-1pt minus 1pt}

Let us start by noting that the four matrix elements appearing in
eq.~(\ref{melkp}) can be obtained in the formal continuum QCD
theory with four quark flavours (qcd4) from (suitable 
Fourier transforms of) the correlators 
\beq
C_{\pm,K\pi}^{({\rm{qcd}}4)}(x,y) = \langle \Phi_{\pi}(x) {\cal
O}^{\pm}_{VV+AA}(0) \Phi^\dagger_{K}(y) \rangle_{\rm conn}
\label{KPI_qcd4mod} 
\eeq
with $K$ and $\pi$ being either $K^0$ and
$\pi^0$ or $K^+$ and $\pi^+$.

\pagebreak[3]

Always at a formal level, we need to consider again the $(4s10v)$
model that was introduced in section~\ref{sec:CFVQ}.
We are interested now in the correlators 
\beq 
C_{\pm , K\pi}^{(4s10v)}(x,y) = \langle
\Phi_{\pi}(x) {\cal Q}^{\pm}_{VV+AA}(0) \Phi^\dagger_{K}(y)
\rangle_{\rm conn}\, , \label{KPI_4s10vmod}
\eeq 
where the meson interpolating operators $\Phi_K$ and $\Phi_\pi$ are made out
exclusively of the valence quarks $u$, $d$ and $s$, while the
operators ${\cal Q}^{\pm}_{VV+AA}$ are given by 
\beqn
{\cal
Q}^{\pm}_{VV+AA} &=& {\cal Q}^{\pm \, [0]}_{VV+AA}
 + {\cal Q}^{\pm \, [1]}_{VV+AA}
 - {1 \over 2} {\cal Q}^{\pm \, [2]}_{VV+AA}
 - {1 \over 2} {\cal Q}^{\pm \, [3]}_{VV+AA} \, ,
\label{Q_COMB} \\
{\cal Q}^{\pm
\,[k]}_{VV+AA} &=& \half \Big[ (\bar{s}\gamma_\mu u^{[k]} )
(\bar{u}^{[k]} \gamma_\mu d ) \pm (\bar{s}\gamma_\mu d )
(\bar{u}^{[k]} \gamma_\mu u^{[k]} ) \Big]  -  \half \Big[
u^{[k]} \leftrightarrow c^{[k]} \Big] + 
\nonumber \\&& 
+ \half \Big[ (\bar{s}\gamma_\mu\gamma_5 u^{[k]}
) (\bar{u}^{[k]} \gamma_\mu\gamma_5 d) \pm
(\bar{s}\gamma_\mu\gamma_5 d ) (\bar{u}^{[k]} \gamma_\mu\gamma_5
u^{[k]} ) \Big]  -  \half \Big[ u^{[k]} \leftrightarrow
c^{[k]}\Big]\, .\qquad
\label{Q_pm_VVAA_l} 
\eeqn 
As in the previous
section, if the two $(4s10v)$ and $({\rm qcd}4)$ models are
regularized in the same way (say, using GW fermions),
a straightforward application of Wick theorem leads to the
equalities
\beq C_{\pm , K\pi}^{(4s10v)}(x,y) =
C_{\pm,K\pi}^{({\rm{qcd}}4)}(x,y) \, , \label{CpmKPIC_EQUAL}
\eeq
valid at any finite value of the lattice spacing, provided bare (and
thus also renormalized) quark masses of the same flavour are given
identical numerical values in the two theories.
Eqs.~(\ref{CpmKPIC_EQUAL}) in particular imply that the
renormalization constants of the operators ${\cal{Q}}^{\pm}_{VV+AA}$
in $(4s10v)$ can be taken equal to those of the operators ${\cal
O}^{\pm}_{VV+AA}$ in $({\rm qcd}4)$. The same equality holds for the
kaon and pion interpolating fields. With this natural choice of
renormalization conditions eqs.~(\ref{CpmKPIC_EQUAL}) are valid for
the corresponding renormalized correlators and carry over to the
continuum limit.

\subsection{A convenient lattice discretization of the
$C_{\pm ,K\pi}^{(4s10v)}$ correlators}
\label{sec:CLD}

For the purpose of getting rid of all mixing problems in the
process of computing the matrix elements~(\ref{melkp}), we will
have to regularize the $(4s10v)$ model in way which is slightly
different from that we followed in section~\ref{sec:CLDC}. Namely,
we now have to take
\beqn 
& -r_s = r_d \, ,  
\nonumber \\
& r_u = -r_{u'} = r_{u''} =  -r_{u'''} = r_c = -r_{c'} = r_{c''} =
-r_{c'''} \, . \label{RF_KtoPI} 
\eeqn 
and fix the relation between the two sets of Wilson parameters in
eq.~(\ref{RF_KtoPI}) by setting\footnote{This choice is made for
convenience, i.e.\ to directly exploit the results of the previous
section in the argument we will give in section~\ref{sec:PAM} and also
to have, for $m_d=m_u$, exact isospin symmetry in the $u$ and $d$
valence quark sector. However, for the purpose of canceling unwanted
mixings only the condition $|r_u|=|r_d|$ is necessary (see
appendix~\ref{sec:APPB}).}
\beq 
r_u= r_d\, .
\label{ncud} 
\eeq 
This lattice regularized model will be denoted in the following as
$(4s10v)^{\rm {L}*}_{tm}$. We stress that it differs from the lattice
model $(4s10v)^{\rm {L}}_{tm}$ of section~\ref{sec:CLDC} only by the
sign of $r_s$ relative to $r_d=r_u$ (and to all the remaining valence
Wilson parameters). Naturally, in order to obtain from the
$(4s10v)^{\rm {L}*}_{tm}$ model correlators that in the continuum
limit are equivalent to those of (qcd4), the quark mass
renormalization
conditions~(\ref{MUDCS_MATCH_LAT1})--(\ref{MUDCS_MATCH_LAT2}) have to
be imposed. As in section~\ref{sec:BKLATT} such conditions entail
eq.~(\ref{MUC_BARE_EQ}).

In practice to extract the matrix elements~(\ref{melkp}),
besides the three-point correlators (\ref{KPI_4s10vmod}), 
it will be necessary to evaluate in the $(4s10v)^{\rm {L}*}_{tm}$ 
model also two-point correlators of the type 
\beq
\langle
\Phi_{\pi}(x) \Phi^\dagger_{\pi}(x') \rangle_{\rm conn}\, ,
\qquad \langle \Phi_{K}(x) \Phi^\dagger_{K}(x') \rangle \, .
\label{aux_KPI_corr} 
\eeq
Suitable spatial Fourier transforms of these two- and three-point
correlation functions contain all the information which is necessary
in order to extract the desired $K\to\pi$ matrix element. The
correlators~(\ref{KPI_4s10vmod}) and~(\ref{aux_KPI_corr}) can be
O($a$) improved by taking appropriate $WA$'s or by averaging over
amplitudes computed with opposite signs of the $K$ and $\pi$
three-momenta.

Analogously to the $K\to\pi\pi$ case previously discussed, also here
one could eliminate the valence quark species $u'''$ and $c'''$ and
accordingly modify the expression of the operator~(\ref{Q_COMB}) with
the substitution
\beq
{\cal Q}^{\pm}_{VV+AA} \to {\cal Q}^{\pm \,
[0]}_{VV+AA}+{1 \over 2} {\cal Q}^{\pm \, [1]}_{VAVAA} - {1 \over
2} {\cal Q}^{\pm \, [2]}_{VV+AA}\, .
\label{REPTTT}
\eeq

\subsection{The renormalization properties of
${\cal Q}^{\pm}_{VV+AA}$}
\label{sec:PAM}

Proving that in the $(4s10v)^{\rm {L}*}_{tm}$ model the operator
${\cal Q}^{\pm}_{VV+AA}$ (eq.~(\ref{Q_COMB})) does not mix with any
other operator is immediate from the results of
section~\ref{sec:NOMIX_O_VAAV}. It suffices, in fact, to perform in
the functional integrals that defines the relevant correlation
functions of the $(4s10v)^{\rm {L}*}_{tm}$ model the change of
fermionic integration variables induced by the transformation
${\cal{R}}_{5s}$ (see eq.~(\ref{R5})). The reason is that under this
transformation 1) the action of the $(4s10v)^{\rm {L}*}_{tm}$ model
turns into that of the $(4s10v)^{\rm {L}}_{tm}$ model except for a
sign inversion in front of the $s$ quark mass term; 2) the operator
${\cal Q}^{\pm}_{VV+AA}$ is transformed into the operator ${\cal
Q}^{\pm}_{VA+AV}$. At this point we are exactly in the situation we
have discussed in section~\ref{sec:DIUM2}, except that we have to
carry out the transformation ${\cal{R}}_{5s}$ on the list of
operators~(\ref{MIX3+3}) to~(\ref{MIX5+4}) and invert the sign of
$m_s$ in all formulae we previously got. Thus, the pattern of
eqs.~(\ref{MIX3+3}) to~(\ref{MIX5+4}) gets modified as follows
\beqn
&& (m_c - m_u) (m_c + m_u) (m_s + m_d)
\hphantom{(m_s - m_d)}\qquad \bar{s}  d \, ,
\label{MIX3+3D} \\
&& (m_c - m_u) (m_c + m_u) (m_s + m_d) (m_s - m_d)
\qquad \bar{s}\gamma_5 d \, ,
\label{MIX3+4D} \\
&& (m_c - m_u) (m_c + m_u) (m_s + m_d)
\hphantom{(m_s - m_d)}\qquad \bar{s}  \sigma \!\cdot\! F d \, , 
\label{MIX5+3D}\\
&& (m_c - m_u) (m_c + m_u) (m_s + m_d) (m_s - m_d) 
\qquad \bar{s} \sigma
\!\cdot\! F \gamma_5 d \, . \qquad
\label{MIX5+4D}
\eeqn
The mass factors appearing in eqs.~(\ref{MIX3+4D}) and~(\ref{MIX5+4D})
in front of the operators $\bar{s}\gamma_5 d$ and $\bar{s}\gamma_5
\sigma\!\cdot\!Fd$ depend quadratically on $m_s$, so they are left
unchanged. Furthermore, the whole line of arguments we gave before to
prove that ${\cal Q}^{\pm}_{VA+AV}$ could not mix with operators of
dimension six in the $(4s10v)^{\rm {L}}_{tm}$ model can be applied to
infer the same property for the operators ${\cal Q}^{\pm}_{VV+AA}$ in
the $(4s10v)^{\rm {L}*}_{tm}$ model, as dimensionless mixing
coefficients are independent of the quark mass (in a mass independent
renormalization scheme). This concludes our proof.

The net result of this analysis is that the situation is just like in
continuum massive QCD. Only the particular combination appearing in
eq.~(\ref{MIX3+3D}) needs a (logarithmically) divergent coefficient,
were this subtraction required.  This term, however, will not
contribute to the $K\to\pi$ form factor that is related through SPT's
to the $K\to\pi\pi$ amplitude, because SPT's have to be understood as
relations between amplitudes evaluated in the chiral limit.

Concerning the renormalization constants of the operators ${\cal
Q}^{\pm}_{VV+AA}$ in the model $(4s10v)^{\rm {L}*}_{tm}$, it is clear
that they are equal to those of the operators ${\cal Q}^{\pm}_{VA+AV}$
in the model $(4s10v)^{\rm {L}}_{tm}$ (see comments on the evaluation
of $Z^{\pm}_{VA+AV}$ at the end of section~\ref{sec:CLDC}).  The
reason is simply that, as pointed out before, the change of fermionic
integration variables induced by the transformation ${\cal{R}}_{5s}$
exactly maps the two theories and the two operators one into the
other.

\section{Concluding remarks}
\label{sec:CONCL}

Our present knowledge of QCD offers a deep understanding of strong
interaction physics with two noticeable exceptions. One is the lack of
a convincing solution of the strong-CP problem~\cite{SCPP}, the second
is the difficulty of finding a physically sound explanation of the
anomalously large value of the ${\rm R}(K\to\pi\pi)$ (eq.~(\ref{DELI})).

Clarification of anyone of these two issues may have a significant
impact on the structure of any forthcoming unified theory comprising
the Standard Model, or give hints about possible signals of new
physics already at low energy, though it is likely that the strong
CP-problem cannot have a solution within QCD~\cite{PQ}, but requires
its embedding in a larger theory.\footnote{See, however, the papers in
ref.~\cite{WMS} for a possible solution within QCD itself.}

In this perspective the results of the present paper look rather
interesting, because, when put together with those of
refs.~\cite{FR1,FR2}, set the basis for a workable framework, where a
first principle computation of the matrix elements of the
CP-conserving $\Delta S=1$ and $\Delta S=2$ effective weak hamiltonian
at realistic values of the light quark masses appears to be feasible
with the computer resources that are or will be soon available. A more
precise estimate of the requested computational effort will be
possible as soon as the practical issues concerning the magnitude of
residual cutoff effects, in particular for light quarks, and possible
lattice phase transitions in unquenched simulations (see the discussion
below) will be clarified.

Independently of these practical issues, the use of maximally twisted
Wilson (or rather \"Osterwalder-Seiler) fermions, according to the
strategy advocated in this paper, brings about conceptual and
qualitative progresses which can be summarized as follows:
\begin{enumerate}
\item no zero modes of the Wilson-Dirac operator at non-vanishing
      quark masses (in particular no exceptional configurations); 
\item a positive definite fermionic determinant, even for mass
      non-degenerate quarks, in unquenched studies with an even number
      of dynamical flavours (in this paper we focused on the
      phenomenologically relevant case of four flavours);
\item no contribution from wrong chirality or opposite parity mixing
      for the lattice operators entering the computation of the weak
      matrix elements of interest;
\item automatic (or easily obtainable) O($a$) improvement of all relevant 
      physical quantities.
\end{enumerate}

Although the method we propose is rather general, in this paper for
definiteness it has been applied only to the (important) instance of
the evaluation of the $B_K$ parameter entering the $K^0$--$\bar{K}^0$
mixing and the $K\to\pi\pi$, or $K\to\pi$, amplitude. In the lattice
computational framework we propose
\begin{itemize}
\item the $B_K$ parameter can be hopefully computed including sea
quark effects, with a sufficiently small error, so as to significantly
reduce one of the largest theoretical uncertainty in the phenomenology
of the unitarity triangle~\cite{BURAS_REV};
\item the yet elusive $\Delta I=1/2$ signal does not appear to be
anymore buried underneath a numerically overwhelming, divergent
subtraction (as was the case for Wilson fermions).
\end{itemize}
These results are obtained at a surprisingly low price. Sea quarks
need to be introduced in pairs and a somewhat exotic pattern of
valence quarks has to be used in the construction of the effective
weak hamiltonian lattice operator.

From several viewpoints the situation is thus very much like with GW
fermions, where chiral symmetry is exact and cut-off effects only show
up at O($a^2$). The non-negligible gain that we get with the strategy
we are advocating here is that the computational burden appears to be
significantly reduced with respect to simulations involving GW
fermions. This feature makes our method potentially suited for the
study of systems with a physical linear size of a few fermi, while
including sea quark effects.  In comparison to approaches based on
untwisted Wilson fermions (which are, however, plagued with
complicated operator mixing and the problem of ``exceptional
configurations''), the method we propose possibly requires a little
more extra computing time, owing to the need of computing certain
valence quark propagators for opposite values of the Wilson parameter.
This overhead is more than compensated at small quark masses by the
expected improvement in the performance of linear solvers, due to the
protection against spurious quark zero modes that is guaranteed by
twisting.

\subsection{Critical mass and O($a$) improvement}

It should be noted that our work is not based on any special
definition of the critical mass for sea and valence quarks (see
eqs.~(\ref{CRMS})--(\ref{CRMV})). The method for O($a$) improvement
discussed in section~\ref{sec:RCI}, which is a straightforward
generalization of the approach presented in ref.~\cite{FR1}, relies
only on the property that $f_{\rm cr}(r_1;r_2,r_3)$ is an odd function
of $r_1$, implying that the sea and valence critical masses,
$M_{\rm{cr}}(r_{\ell,h})$ and $M_{\rm{cr}}(r_f)$ are in turn odd in
$r_{\ell,h}$ and $r_f$, respectively.  The argument showing that
$f_{\rm cr}$ is odd in $r_1$ we give here (in section~\ref{sec:WTIS}
and appendix~\ref{sec:APPA}) is analogous to that presented in
ref.~\cite{FR1} in the case of generic (chirally twisted) Wilson
fermions.

The question of the $r$-parity properties of $M_{\rm{cr}}(r)$ was
taken up in refs.~\cite{AOBA,BAER}. Clearly this issue could have an
impact on the question of O($a$) improvement as we have discussed it
in the present paper and in ref.~\cite{FR1}. Actually, it turns out
that $M_{\rm{cr}}(r)$ can and should always be taken odd in $r$.

The reason is simply that any counterterm to be included in the
process of renormalizing the theory should not break any of the
symmetries the classical action enjoys prior to
renormalization. Indeed, the manifold of the possible solutions for
$M_{\rm cr}$, defined by whichever condition one wishes to take, has
the property of being reflected around zero under the transformation
$r\to -r$ (see e.g.\ ref.~\cite{F_PROC} for a discussion of this
point).  Thus it would be unwise and not very useful to violate
the (spurionic) symmetries of the theory (in particular ${\cal
P}\times (R\rightarrow -R)$) by having a critical mass that, for
instance, to some O($a^p$) with $p>1$, is not odd in $r$.

Apart from this observation, it is important to remark that the whole
issue of O($a$) improvement, as it is discussed in ref.~\cite{FR1} and
in this paper, assumes that as $a\to 0$ the counting of powers of $a$
is done by taking the renormalized quark masses fixed.  We did not
address the interesting question of what happens when the quark mass
is taken to be a quantity of O($a$) or O($a^2$), as often done in
papers dealing with lattice chiral perturbation
theory~\cite{AOBA,BAER,MSS,SHSI,MSW}.

\subsection{Tests and numerical issues}

The numerical effectiveness of the whole approach based on maximally
twisted Wilson quarks has to be carefully checked, especially for
small quark masses and when taking into account the effects of sea
quarks.

There are already encouraging results, coming from the quenched
simulations of ref.~\cite{KIAC} and the analytic work of
ref.~\cite{MSS}, indicating that the scaling behaviour of the theory
is indeed smoother and flatter than with standard or clover-improved
Wilson fermions. An extension of the scaling tests of ref.~\cite{KIAC}
to several lower values of the pseudoscalar meson mass, down to less
than 300\,MeV, is currently in progress~\cite{SHTALK}.

Furthermore, recently, an exploratory study of the effects of
unquenching in simulations of tm-LQCD with two mass degenerate quarks
on rather coarse lattices of size $8^3\times 16$, $12^3\times 24$ and
$16^3\times 32$ at $\beta=5.2$ (using the standard Wilson plaquette
action) was presented~\cite{10AUTH}. In this work the PCAC quark mass,
$m_\chi^{PCAC}$ (see eq.~(12) of ref.~\cite{10AUTH}), the mass of the
pion and the plaquette expectation value have been measured at
$am_\ell=0.01$,\footnote{In the notation of ref.~\cite{10AUTH} $m_\ell
\equiv \mu$.} as functions of the untwisted mass,
$m_0$.\footnote{Introducing $m_0$ in the action~(\ref{FULACT}) amounts
to replacing there $W_{\rm{cr}}$ with $W_{\rm{cr}}-M_{\rm{cr}}+m_0$.}

Results show features of the kind suggested by
refs.~\cite{AOKI,SHSI,MSW} in the framework of chiral perturbation
theory, when the order of magnitude inequality
$a^2\Lambda^2_{\rm{QCD}}\ll am_\ell$ is not
satisfied~\cite{FR1}. Metastabilities are detected, signalled by the
behaviour of the plaquette expectation value which is seen to undergo
hysteresis cycles as $m_0$ is driven through $M_{\rm{cr}}$. In this
situation gauge configurations cannot be considered as properly
thermalized. Metastable states associated with values of
$m_\chi^{PCAC}$ of different sign have been identified and argued to
stem from the effective lack of criticality of the subtracted Wilson
term and the associated existence of non-zero values for the parity
and flavour breaking quark condensates.  As a consequence, one does
not yet observe the expected vanishing of $m_\chi^{PCAC}$ at a
critical value of $m_0$ (providing a natural definition of
$M_{\rm{cr}}$). It is, however, reassuring to see that the squared
pion mass appears to be roughly proportional to $\sqrt{(Z_A
m_\chi^{PCAC})^2 + m_\ell^2}$ in both the observed metastable
branches.

On the basis of the present understanding~\cite{MSW} of the
metastability phenomena discussed above, we expect them to get weaker
and weaker and eventually disappear as the lattice spacing is
decreased, making thus possible the evaluation of physical quantities
with controlled O($a^2$) cut-off effects. Preliminary data~\cite{CU}
coming from simulations in progress at $\beta=5.3$ seem to be
consistent with such an expectation.

\subsection{Outlook}

The idea of using a hybrid approach where valence and sea quarks are
regularized in different fashion was already put forward in the
literature (see, for instance, ref.~\cite{SHOR} and references
therein). Along this line a winning strategy, where all the nice
properties we have listed above remain valid, could be to use GW
fermions as valence quarks and maximally twisted quark pairs in
loops. With this choice one could ease the problem related to the need
of working at relatively large values of $\beta$ (and correspondingly
large lattices) if the valence quark mass is small, and obtain a
computational framework, where CPU times for inverting the overlap
Dirac operator and stochastically taking into account the fermion
determinant are comparable.

We end by recalling that the tm-LQCD action can also be
employed~\cite{TMLQCD,tm-ALPHA} in the framework of the Schr\"odinger
functional formalism~\cite{SFALPHA}. Results about the O($a$)
improvement remain valid for quantities that are independent of the
boundary conditions, like on-shell matrix elements or masses.
However, if one wishes to improve the whole Schr\"odinger functional,
which is a necessary step in order to get cancellation of O($a$)
cut-off effects in the evaluation of, for instance, the running gauge
coupling, quark masses and renormalization constants, the way of
extending the $WA$ method is not completely obvious and will be the
object of a forthcoming publication~\cite{FR4}.

\acknowledgments

We thank D.~Be\'cirevi\'c, M.~Della~Morte, L.~Giusti, C.~Hoelbling,
K.~Jansen, L.~Lellouch, C.~Pena, M.~Testa and A.~Vladikas for
discussions. One of us (G.C.R.) wishes to thank the Humboldt
Foundation for financial support.

\appendix

\section{Renormalizability of the lattice model~(\protect\ref{FULACT})}
\label{sec:APPA}

We want to show in this appendix that the action of the lattice
model~(\ref{FULACT}) contains all the terms of dimension not
larger than four, allowed by the action symmetries, that are
required for correlators with no ghost fields be renormalizable.
In the course of this discussion we also prove 1) the results
contained in eqs.~(\ref{SEAMQRE})--(\ref{VALMQRE}), 2) the
property that a unique function, $f_{\rm cr}(r;r_\ell,r_h)$,
determines the critical mass of OS valence and twisted sea quarks
(see eqs.~(\ref{CRMS})--(\ref{CRMV})) and 3) the relation
$Z_m(r_f)=Z_P(r_f)^{-1}$ for $r_f = r_\ell$ or $r_f = r_h$.

We start by noticing that completely standard arguments, based on 
dimensionality and invariance of the action~(\ref{FULACT}) under lattice 
gauge transformations, translations, $H(4)$ rotations and charge conjugation,
\beq 
{\cal{C}}:\left
\{\begin{array}{rclrcll}
U_\mu(x)&\rightarrow& U_\mu^*(x) & 
\\
\psi_p(x)&\rightarrow& i \gamma_0\gamma_2
\bar\psi^T_p(x) \, ,
\qquad& 
\bar{\psi}_p(x) &\rightarrow& -\psi^T_p(x) i \gamma_0 \gamma_2 \, , \qquad& p=\ell,h 
\\
q_f(x)&\rightarrow& i \gamma_0\gamma_2 \bar{q}^T_f(x) \, ,
\qquad&
\bar{q}_f(x) &\rightarrow& -q^T_f(x) i \gamma_0 \gamma_2\, ,\qquad&\hbox{all } f{\rm's} 
\\
\phi_f(x) &\rightarrow& \multicolumn{4}{l}{i \gamma_0 \gamma_2
(\phi^\dagger_f)^T(x) = i \gamma_0 \gamma_2 \phi^*_f(x) \, ,}
\qquad&\hbox{all } f{\rm's} 
\end{array}\right . \, , 
\label{CHARGE}
\eeq
imply that the only relevant action density terms are, besides
gauge invariant gluonic operators ($FF$ and $F\widetilde{F}$), either
fermionic bilinears of the type $\bar{\psi}_p \dots \psi_p$
and $\bar{q}_f \dots q_f$ or ghost bilinears of the form
$\phi_f^\dagger \dots \phi_f$.

\subsection{Sea quark sector}
\label{sec:SQS}

The renormalization of correlation functions with only gluons and
$\psi_{\ell,h}$, $\bar{\psi}_{\ell, h}$ fields proceeds as discussed
in ref.~\cite{FR2}. In fact, for this class of correlation functions
the contributions coming from functionally integrating valence quark and
ghost fields completely cancel against each other, while the
symmetries discussed in ref.~\cite{FR2} are straightforwardly extended
from one to two pairs of maximally twisted quarks.\footnote{In
particular the symmetry operations that do not leave inert the gauge
field must be performed on both quark pairs simultaneously. Important
examples of these symmetries are
$$ 
{\cal P}_{\pi/2}^1 \times (m_l \to -m_l) \times (m_h \to -m_h) \, ,
\qquad
{\cal P}_{F}^2 \times (\epsilon_l \to -\epsilon_l) \times
(\epsilon_h \to -\epsilon_h) \, ,
$$
with ${\cal P}_{\pi/2}^1$ and ${\cal P}_{F}^2$ representing the
obvious extensions of the operations in eqs.~(62)--(63) of
ref.~\cite{FR2} to the theory with two quark flavour pairs.}

Furthermore mixed sea quark action terms of the form $\bar{\psi}_\ell
\dots \psi_h$ or $\bar{\psi}_h \dots \psi_\ell$ can not arise, as it
trivially follows from the invariance of the action under the
independent $\U(1)$ vector transformations
\beq
{\cal I}_p \; : \psi_p(x) \to e^{i\theta_p} \psi_p(x) \, ,
\qquad 
\bar{\psi}_p(x) \to \bar{\psi}_p(x) e^{-i\theta_p} \, ,
\qquad p=\ell, h \, .
\label{FIXPAIROT}
\eeq

The important equation~(\ref{SEAMQRE}) can be derived from the
analysis of the chiral WTI's of the lattice model~(\ref{FULACT})
involving correlators with the insertion of $\psi_{\ell,h}$ and
$\bar{\psi}_{\ell,h}$ fields only. Since valence quarks are
totally inert, the structure of these WTI's is identical to that
reported in ref.~\cite{FR2} for maximally twisted lattice QCD with
one pair of mass non-degenerate quarks. For the reader convenience
we report here the WTI's from which the renormalization properties
of the quark masses most straightforwardly follow. 
If $\widehat{O}(y)$ is a renormalized 
(multi-local) operator, one gets, among others, the two WTI's ($x\neq y$) 
\beqn
\left\langle\Big{(}{\partial}^\star_\mu\hat{A}^{2p}_\mu(x)
-2\hat{m}_p\hat{P}^{2p}(x)\Big{)}{\hat{O}}(y)\right\>\Big{|}_{(R,M)}
&=&{\mbox{O}}(a)\, ,
\label{A2}\\
\left\langle\Big{(}{\partial}^\star_\mu\hat{A}^{3p}_\mu(x)
-2\hat{m}_p \hat{P}^{3p}(x)
+ \hat{\epsilon}_p\hat{P}^{0p}(x)\Big{)}
{\hat{O}}(y)\right\>\Big{|}_{(R,M)}&=&{\mbox{O}}(a)
\label{A3}\, , \qquad
\eeqn
where $p$ runs over all possible sea quark pairs present in the
theory, $p=\ell,h$. In eqs.~(\ref{A2})--(\ref{A3}) we have used
the definitions 
\beqn
\hat{A}_\mu^{2p}&=&Z_{V}\,\bar\psi_p\gamma_\mu\gamma_5\frac{\tau_2}{2}\psi_p
\, ,\qquad
\hat{A}_\mu^{3p}=Z_V\,\bar\psi_p\gamma_\mu
\gamma_5\frac{\tau_3}{2}\psi_p\, , 
\label{D2}\\
\hat{P}^{3p}&=&Z_{P}\,\bar\psi_p\frac{\tau_3}{2}\gamma_5\psi_p\,, \qquad\hspace{12pt}
\hat{P}^{0p}=Z_{S}\,\bar\psi_p\gamma_5\psi_p\, ,
\label{D4}
\eeqn
and ${\partial}^\star$ is the backward
derivative. The above formulae imply the relations 
\beqn
\hat{m}_p^{+}&=&\hat{m}_p+\hat{\epsilon}_p=Z_P^{-1}m_p+Z_S^{-1}\epsilon_p\, ,
\label{RMP}\\
\hat{m}_p^{-}&=&\hat{m}_p-\hat{\epsilon}_p=Z_P^{-1}m_p-Z_S^{-1}\epsilon_p \, ,
\label{RMM} 
\eeqn
for $p=\ell$ and $p=h$.

The renormalization constants $Z_V$, $Z_P$ and $Z_S$ are even
functions of the Wilson parameter of the quark pair we focus on and
the $r$ parameters of all the other sea quark pairs appearing in
virtual loops, $Z_I(r_p)=Z_I(r_p;r_\ell,r_h)$, $I=V,P,S,\ldots$  The
renormalization constants $Z_I(r_p)$ coincide with the corresponding
ones of the standard Wilson theory with four flavours, if a mass
independent renormalization scheme is employed.

\subsection{Valence quark sector}
\label{sec:VQS}

Let us now come to the proof that the valence quark sector of the
action~(\ref{FULACT}) has the most general renormalizable form
compatible with the symmetries of the model.

Due to valence flavour conservation, each valence flavour $f$
renormalizes independently from all the others. Moreover one verifies
the validity of the following statements.
\begin{enumerate}
\item The $d=4$ operator $\bar
      q_f\gamma_5\gamma\cdot\widetilde{\nabla} q_f$ is forbidden by
      ${\cal{C}}$;
\item The $d=3$ operator $i \bar{q}_f \gamma_5 q_f$ can appear
with a coefficient (to be identified with the critical mass),
proportional to $1/a$ and odd in $r_f$. Such a term is allowed by
the spurionic invariance ${\cal P}_5 \times (M \to - M)$, where
\beq 
{\cal P}_{5} : 
\left \{\begin{array}{rclrcll} 
U_0(x)&\rightarrow& U_0(x_P)\, ,
&
U_k(x)&\rightarrow& U_k^{\dagger}(x_P-a\hat{k})\,,
\quad&
k=1,2,3\\
\psi_p(x)&\rightarrow& \gamma_5 \gamma_0 \psi_p(x_P)\, ,
\qquad&
\bar{\psi}_p(x)&\rightarrow& - \bar{\psi}_p(x_P) \gamma_0 \gamma_5  \, , 
\quad &
p=\ell,h \\
q_f(x)&\rightarrow& \gamma_5 \gamma_0 q_f(x_P) \, ,
&
\bar{q}_f(x)&\rightarrow& - \bar{q}_f(x_P)\gamma_0\gamma_5\, ,
\quad&
\hbox{all } f{\rm's}\\
\phi_f(x) &\rightarrow& \gamma_5 \gamma_0 \phi_f(x_P)\, ,
&&& 
\quad&
\hbox{all } f{\rm's} \, ,
\end{array}\right . 
\label{P5} 
\eeq
as such symmetry only excludes coefficients odd in $m_f$. The
spurionic invariance ${\cal R}_{5f}^{\rm sp}$ (see
eq.~(\ref{R5SPVAL})) fixes the $r_f$-parity of the coefficient to be
negative. We recall that no odd powers of mass can arise from the sea
quark determinant, as the latter is even in the sea quark mass
parameters (as well as in the sea Wilson parameters). This proves the
statement, made in the text, that the critical mass $M_{\rm cr}(r_f) =
a^{-1}f_{\rm cr} (r_f;r_\ell,r_h)$ is an odd function of $r_f$.
\item The $d=3$ operator $\bar{q}_f q_f$ is allowed, owing to
      the spurionic invariance ${\cal P}_5 \times (M \to - M)$, only
      if multiplied by an odd power of $m _f$ (again no odd power
      of $m_{\ell,h}$ or $\epsilon_{\ell,h}$ can arise from the sea
      quark determinant). Hence a term like $a^{-1}\bar{q}_f q_f$
      is ruled out and only the term $m_f \bar{q}_f q_f$ needs
      be included in the action. This implies the multiplicative
      renormalizability of the valence quark mass parameter, $m_f$,
      see eq.~(\ref{VALMQRE}).\footnote{The analysis of chiral WTI's with 
      insertions of valence quark operators leads to the same result.}
      As remarked before, the renormalization constant of $m_f$,
      called $Z_m(r_f)$ in the text, must be even in $r_f$, for 
      consistency with the spurionic invariance ${\cal R}_{5f}^{\rm sp}$.
\item The reality properties of the coefficients of all the quark
      bilinears are fixed by site/link reflection invariance,
      $\Theta_{s/l}$. The action of $\Theta_{s/l}$ on gauge and sea
      quark fields has been specified in eqs.~(58)--(59) of
      ref.~\cite{FR2}. Valence quarks, $q_f$, and ghosts, $\phi_f$,
      transform under $\Theta_{s/l}$ precisely as each member of the
      sea quark pair field, $\psi_p$, while $\bar{q}_f$ transforms as
      the corresponding antiquark field.
\end{enumerate}
Since we are not interested in correlators with ghost field
insertions, we need not discuss in detail the renormalization of the
ghost sector of the model. We only wish to stress again that, as we
already remarked in the text, the parameters, $r_f$ and $m_f$, in the
ghost action~(\ref{VALGACT}) are completely fixed by the requirement
that integration over the ghost field should exactly cancel the
valence fermion determinant.\footnote{Incidentally if one is willing
to consider graded field transformations between valence fermions and
ghosts~\cite{SHSH}, then it turns out that, once the valence fermion
action is fixed by the usual renormalization conditions, the form of
the ghost action is determined by requiring invariance under e.g.\ the
transformation (with super-determinant equal to 1) $ \phi_f(x) \to
q_f(x)$, $q_f(x) \to \phi_f(x)$, $\phi_f^\dagger(x) \to {\rm
sign}(m_f) \bar{q}_f(x)$, $\bar{q}_f(x) \to {\rm sign}(m_f)
\phi_f^\dagger(x)$,  which for each flavour exchanges quark and ghost
fields.}

It is also a trivial observation that the invariance of the
model~(\ref{FULACT}) under independent $\U(1)$ transformations acting
either on single valence flavours (${\cal I}_f$, see
eq.~(\ref{FIXFLAROT})) or individual pairs of sea quarks (${\cal
I}_p$, see eq.~(\ref{FIXPAIROT})) guarantees that no fermion bilinears
with mixed valence-sea quark content can arise.

As for the operator $iF\widetilde{F}$, its presence can be induced
neither by integrating over the
sea sector~\cite{FR2}, because of the invariance under parity
(${\cal P}$, see eq.~(\ref{PAROP})) combined with
$$
\psi_{\ell,h}(x) \to i\tau_3  \psi_{\ell,h}(x)\, ,\qquad
\bar{\psi}_{\ell,h}(x) \to -\bar{\psi}_{\ell,h}(x) i \tau_3\, , 
$$
nor by integrating over the valence sector because the resulting
fermionic determinant is exactly canceled by the ghost
integration.

We note incidentally that the transformation ${\cal P}_{5}$
(see eq.~(\ref{P5})) can be also cast in the form
\beq
{\cal P}_{5} = {\cal P} \times {\cal
R}_{5l} \times {\cal R}_{5h} \times \prod_f {\cal R}_{5f}\, ,
\label{P5_DECOMP}
\eeq
where the transformations ${\cal R}_{5f}$ and ${\cal R}_{5p}$
are defined in eqs.~(\ref{R5})--(\ref{R5SEA}).

\subsection{Sea and valence quark mass renormalization}
\label{SVQMR}

Here we want to establish the fact that within the model~(\ref{FULACT}) 
a unique dimensionless function, $f_{\rm cr}$, determines the critical 
masses in eqs.~(\ref{CRMS})--(\ref{CRMV}) and prove the relation~(\ref{ZMZP}) 
between the quark mass renormalization constants of sea and valence quarks.

For this purpose, let us consider correlators containing either
only the fields $\psi_\ell$ and $\bar{\psi}_\ell$ or only a pair
of valence flavour fields, that we collect in the doublet $Q
\equiv (q_1, q_2)$, and choose
\beq
r_\ell = r_1 = -r_2 \equiv r \, , \qquad m_\ell = m_1 = m_2 \equiv m\, ,
\quad \epsilon_\ell =0 \, . \label{SETUP}
\eeq
In this setting the relation between quark mass renormalization
constants we wish to prove reads
\beq \label{Z_M_REL}
Z_m(r;r,r_h) = Z_P^{-1}(r;r,r_h) \, .
\eeq
It is convenient to consider two pairs of WTI's, one involving
correlators made out of exclusively the fields $\psi_\ell$ and
$\bar{\psi}_\ell$ and the second one involving only the valence fields
$Q$ and $\bar{Q}$. The WTI's of interest are ($x \neq y$)
\beqn
\partial_\mu^* \langle V_\mu^{2 \, \ell}(x) P^{2 \,
\ell}(y)\rangle\Big{|}_{(R,M)}&=& {\rm O}(a) \, , 
\label{VWI_S} \\
\partial_\mu^* \langle A_\mu^{2 \, \ell}(x) P^{2 \,
\ell}(y) \rangle\Big{|}_{(R,M)} &=& 2 m \langle P^{2 \, \ell}(x)
P^{2 \, \ell}(y) \rangle\Big{|}_{(R,M)} + {\rm O}(a)\, ,
\label{AWI_S}
\eeqn
where we have used the definitions
\beqn
V_\mu^{2 \, \ell}(x) &=& \bar{\psi}_\ell(x) \gamma_\mu \frac{\tau_2}{2}
\psi_\ell(x) \, ,
\label{OPV_S}\\
P^{2 \, \ell}(x) &=& \bar{\psi}_\ell(x) \gamma_5 \frac{\tau_2}{2}
\psi_\ell(x) \,,
\label{OPP_S} \\
2\,A_\mu^{2 \, \ell}(x) &=& \bar{\psi}_\ell(x) \gamma_\mu \gamma_5
\frac{\tau_2}{2}U_\mu(x) \psi_\ell(x+a\hat{\mu}) + \bar{\psi}_\ell(x+a\hat{\mu})
\gamma_\mu \gamma_5
\frac{\tau_2}{2} U_\mu^\dagger(x) \psi_\ell(x) -\qquad
\nonumber \\&&
-r \biggl[ \bar{\psi}_\ell(x) \frac{\tau_3}{2} U_\mu(x) \psi_\ell(x+a\hat{\mu})
-\bar{\psi}_\ell(x+a\hat{\mu}) \frac{\tau_3}{2} U_\mu^\dagger(x) \psi_\ell(x)
\biggr] 
\label{OPAC_S}
\eeqn
and the corresponding WTI's with the $\psi_\ell$ and $\bar{\psi}_\ell$
fields replaced by the valence fields $Q$ and $\bar{Q}$, i.e.\
\beqn
\partial_\mu^* \langle V_\mu^{2 \, Q}(x) P^{2 \, Q}(y) \rangle\Big{|}_{(R,M)}
&=& {\rm O}(a) \, , 
\label{VWI_V}\\
\partial_\mu^* \langle
A_\mu^{2 \, Q}(x) P^{2 \, Q}(y) \rangle\Big{|}_{(R,M)} &=& 2 m
\langle P^{2 \, Q}(x) P^{2 \, Q}(y) \rangle\Big{|}_{(R,M)} + {\rm
O}(a) \, , \qquad
\label{AWI_V}
\eeqn
where analogously we have set 
\beqn
V_\mu^{2 \, Q}(x) &=& \bar{Q}(x) \gamma_\mu \frac{\tau_2}{2} Q(x) \, ,
\label{OPV_V}\\
P^{2 \, Q}(x) &=& \bar{Q}(x) \gamma_5 \frac{\tau_2}{2} Q(x) \,,
\label{OPP_V} \\
2\,A_\mu^{2 \, Q}(x) &=& \bar{Q}(x) \gamma_\mu \gamma_5 \frac{\tau_2}{2}
U_\mu(x) Q(x+a\hat{\mu}) + \bar{Q}(x+a\hat{\mu}) \gamma_\mu \gamma_5
\frac{\tau_2}{2} U_\mu^\dagger(x) Q(x) +\qquad
\nonumber \\&& 
+r \biggl[ \bar{Q}(x) \frac{\tau_1}{2} U_\mu(x) Q(x+a\hat{\mu})
   - \bar{Q}(x+a\hat{\mu}) \frac{\tau_1}{2} U_\mu^\dagger(x) Q(x)
   \biggr] \, . 
\label{OPAC_V}
\eeqn 
Imposing the validity of the WTI's~(\ref{VWI_S}) and~(\ref{VWI_V}) is
a way of arriving at a definition of the critical mass for the
$\ell$ight quark pair and the valence quark doublet $Q=(q_1, q_2$),
respectively.  The key observation is that, with the
choice~(\ref{SETUP}) of bare parameters, the l.h.s.\ of
eq.~(\ref{VWI_S}) is exactly equal to the l.h.s.\ of
eq.~(\ref{VWI_V}).  To prove this it is enough to perform a (suitable)
vector-$\tau_2$ rotation of the $\ell$ight quark field pair, so as to
rewrite its action in the basis where the Wilson term is multiplied by
$-i\gamma_5\tau_3$ as is the case of the $Q$-doublet.  Such a
rotation does not alter the expression of the operators~(\ref{OPV_S})
and~(\ref{OPP_S}), but brings the flavour structure of the axial
current~(\ref{OPAC_S}) to that of eq.~(\ref{OPAC_V}).  It thus follows
that the numerical value of $M_{\rm cr}=a^{-1}f_{\rm cr}(r;r,r_h)$
inferred by enforcing the vanishing of the r.h.s.\ of
eq.~(\ref{VWI_S}) is the same that is obtained by enforcing the same
condition on eq.~(\ref{VWI_V}), and viceversa. This proves the
relations~(\ref{CRMS})--(\ref{CRMV}).

The WTI's~(\ref{AWI_S}) and~(\ref{AWI_V}) contain all the relevant
information about the quark mass of the $\ell$ight quark pair and the
valence quark doublet ($q_1, q_2$), respectively.  Moreover they
involve the exactly conserved 1-point split axial
currents~(\ref{OPAC_S}) and~(\ref{OPAC_V}). Precisely for this reason,
the quark mass renormalization factor can be taken equal to the
inverse of the renormalization constant of the relevant pseudo-scalar
densities, which are $P^{2\,\ell}$ (eq.~(\ref{OPP_S})) and $P^{2\,Q}$
(eq.~(\ref{OPP_V})) for the $\ell$ight quark pair and the valence
quark doublet, respectively.  In the text we have denoted the mass
renormalization constant of the $\ell$ight quark pair by
$Z_P^{-1}(r)=Z_P^{-1}(r;r,r_h)$ and that of the valence fields
$Q=(q_1,q_2)$ by $Z_m(r)=Z_m(r;r,r_h)$ (recall that $Z_m$ is even in
$r$ and we have taken $r_1=-r_2 \equiv r$, see eq.~(\ref{SETUP})). At
this point the relation~(\ref{Z_M_REL}) is a consequence of the fact
that, by the same argument we gave before for the vector
WTI's~(\ref{VWI_S}) and~(\ref{VWI_V}), the correlator $\langle P^{2 \,
\ell}(x) P^{2 \, \ell}(y) \rangle |_{(R,M)}$ in eq.~(\ref{AWI_S}) is
equal to the correlator $\langle P^{2 \,Q}(x) P^{2 \, Q}(y) \rangle
|_{(R,M)}$ in eq.~(\ref{AWI_V}).

\section{Isospin and parity breaking effects in
the light quark sector}
\label{sec:APPB}

Although to guarantee the desired renormalization properties of ${\cal
Q}^{\pm}_{VA+AV}$ only the condition $|r_u| = |r_d|$ is actually
necessary, the choice~(\ref{RF_pions_NEW}) is quite convenient as it
preserves (for $m_d = m_u$) the isospin symmetry of the action of the
$u$ and $d$ pair of valence quarks. As a consequence of this symmetry,
the two-pion states $|\pi^+\pi^->$ and $|\pi^0\pi^0>$ cannot mix with
the neutral pion state, $|\pi^0>$, as it would be the case had we
chosen the condition $r_u=-r_d$ instead of~(\ref{RF_pions_NEW}). This
remark is very important in practice, because (recall that parity is
not a symmetry of tm-LQCD) it allows to exclude the presence of the
neutral pion state among the low mass states contributing to the
spectral representation of the correlators $C_{\pm
,\pi^+\pi^-}^{(4s10v)}(x,z,y)$ and $C_{\pm
,\pi^0\pi^0}^{(4s10v)}(x,z,y)$ (see
eqs.~(\ref{K2PIC_modcorr})--(\ref{K2PIN_modcorr})).

It should be noted, in fact, that, if present (as in the case isospin
is broken by the choice $r_u=-r_d$ or by taking $m_d \neq m_u$), the
contribution of the neutral pion may numerically compete with the
relevant isospin and parity preserving contributions, since its mass
is generically lower than the energy of the two pion states we are
interested in. Consistently with the general properties of tm-LCDQ it
can, however, be shown that isospin and parity breaking contributions
to the spectral representation are mere O($a^2$) cut-off effects.  The
latter can be disentangled by standard techniques and removed from the
measured correlators before extracting the relevant O(1)
contributions.

There is an extra problem that should be mentioned if isospin (and
parity) is broken. Namely the fact that the standard interpolating
field for the neutral pion, eq.~(\ref{ANNIH_PFIELD}), mixes with the
identity operator. To avoid this problem one may replace the latter
with the fourth component of the neutral axial current
$$
\Phi_{\pi^0} = \Phi^\dagger_{\pi^0}\to{1\over\sqrt{2}} 
\Big[\bar{u} \gamma_0\gamma_5 u - \bar{d}
\gamma_0\gamma_5 d \Big] \, .
$$

\section{Absence of the mixing of
${\cal Q}^{\pm}_{VA+AV}$ with operators of dimension six}
\label{sec:APPC}

In this appendix we prove that the operators ${\cal Q}^{\pm \,
[k]}_{VA+AV}$ cannot mix with any other operator of dimension six.
Looking at eqs.~(\ref{RF_Kto2PI})--(\ref{RF_pions_NEW}), we see
that just two operators need to be considered, i.e.\ 
\begin{enumerate}
\item ${\cal Q}^{\pm \, [0]}_{VA+AV}$, where valence quark species
have $r_u = r_c = r_d = r_s$;
\item ${\cal Q}^{\pm \, [1]}_{VA+AV}$, where valence quark species
have $r_{u'} = r_{c'} = -r_d = -r_s$.
\end{enumerate}
Indeed the case of ${\cal Q}^{\pm\,[2]}_{VA+AV}$ is identical to the
case of ${\cal Q}^{\pm\,[0]}_{VA+AV}$, and the case of ${\cal
Q}^{\pm\,[3]}_{VA+AV}$ to that of ${\cal Q}^{\pm \, [1]}_{VA+AV}$,
because the quarks that enter have an equal pattern of signs of the
Wilson parameters.\footnote{The arguments that follow would go through
also if we were to replace eq.~(\ref{RF_pions_NEW}) with $r_u = -
r_d$, while maintaining eq.~(\ref{RF_Kto2PI}). In fact, this new
choice for the valence Wilson parameters is related to the one we made
in the text by the transformation ${\rm Ex}(u,u') \times {\rm
Ex}(u'',u''') \times {\rm Ex}(c,c') \times {\rm Ex}(c'',c''')$ which
leaves invariant the form of the operator ${\cal
Q}^{\pm}_{VA+AV}$. See, however, appendix~\ref{sec:APPB}.}

Since we are discussing the mixing of operators of the same dimension, we
can set to zero the masses of the quarks and consider the lattice model
$(4s10v)^{\rm L}_{tm}$ in the chiral limit.

Let us discuss separately the case of ${\cal Q}^{\pm\,[0]}_{VA+AV}$
and ${\cal Q}^{\pm\,[1]}_{VA+AV}$.

\subsection{The operator ${\cal Q}^{\pm \, [0]}_{VA+AV}$}
\label{sec:OP1}

A basis for the operators of dimension six which ${\cal Q}^{\pm \,
[0]}_{VA+AV}$ can mix with is given by~\cite{GAS} 
\beqn & {\cal
Q}^{+ \, [0]}_{Y} \, , \qquad {\cal Q}^{- \, [0]}_{Y} \,
,\nonumber \\
& Y = VA+AV,\,\, VA-AV,\,\, SP+PS,\,\, SP-PS,\,\, T\widetilde{T},
\nonumber \\& \qquad VV+AA,\,\, VV-AA,\,\, SS+PP,\,\,
SS-PP,\,\, TT \, . \label{OPLISTK2PI} 
\eeqn 
The notation $VA+AV$, $VA-AV$, \dots, $TT$ refers to the Dirac
structure of the four-fermion operators and is self-explanatory.
For instance, with $\sigma_{\mu\nu}\equiv
i[\gamma_\mu,\gamma_\nu]/2$, we have 
\beqn 
{\cal Q}^{\pm \, [0]}_{VV+AA} &=& \half \Big[ (\bar{s}\gamma_\mu u)
(\bar{u} \gamma_\mu d) \pm (\bar{s}\gamma_\mu d) (\bar{u}
\gamma_\mu u) \Big]  -  \half [ u \leftrightarrow c ] +
\nonumber \\&& 
+ \half \Big[ (\bar{s}\gamma_\mu \gamma_5 u) (\bar{u}
\gamma_\mu\gamma_5 d) \pm (\bar{s}\gamma_\mu\gamma_5 d) (\bar{u}
\gamma_\mu\gamma_5 u) \Big]  -  \half [ u \leftrightarrow c
] \, ,\qquad
\label{Q_pm_VVAA_0} \\
 {\cal Q}^{\pm \, [0]}_{TT} &=&
\half \Big[ (\bar{s} \sigma_{\mu\nu} u) (\bar{u} \sigma_{\mu\nu}
d) \pm (\bar{s} \sigma_{\mu\nu} d) (\bar{u} \sigma_{\mu\nu} u)
\Big]  -  \half [ u \leftrightarrow c ] \,.
\label{Q_pm_TT_0} 
\eeqn 
The key observation is that, owing to the equality $r_u = r_c = r_d =
r_s$ of the Wilson parameters of the quarks appearing in ${\cal
Q}^{\pm\,[0]}_{Y}$, the part of the lattice action relative to these
quarks (the standard $u$, $d$, $s$ and $c$ quarks) admits an exact
$\SU(4)$ flavour symmetry, which prevents the operators of the type
$(+)$ to mix with any of the $(-)$ counterparts, as they belong to
different irreducible representations of the $\SU(4)$ flavour
group. The operators in~(\ref{OPLISTK2PI}) can be thus viewed to yield two
separate bases, one for the $(+)$ sector and another one for the $(-)$
sector, containing the candidates for the mixing with ${\cal Q}^{+
\,[0]}_{VA+AV}$ and ${\cal Q}^{- \, [0]}_{VA+AV}$, respectively.

Notice that conservation of each separate valence quark species
forbids the mixing of the operators ${\cal Q}^{\pm\,[0]}_{VA+AV}$ with
the operators ${\cal Q}^{\pm\,[k]}_{Y}$, $k=1,2,3$, as they have
different flavour content.\footnote{Of course, away from the chiral
limit each of the ${\cal Q}^{\pm\,[k]}_{Y}$, $k=0,1,2,3$ can mix with
$\Delta s = -\Delta d =1$ two-quark operators having the same unbroken
quantum numbers, as we discussed in the text.}

For brevity, we will deal ``in parallel'' with ${\cal
Q}^{+\,[0]}_{VA+AV}$ and ${\cal Q}^{-\,[0]}_{VA+AV}$. One checks that
the invariance of the action $(4s10v)^{\rm L}_{tm}$ under ${\cal P}_5$
(see eq.~(\ref{P5})) rules out the mixing of ${\cal
Q}^{\pm\,[0]}_{VA+AV}$ with the operators of the form ${\cal Q}^{\pm
\, [0]}_{W}$, having the Dirac structure $W=VV+AA, VV-AA, SS+PP,
SS-PP, TT$. In fact, these operators are even under ${\cal P}_5$,
while ${\cal Q}^{\pm \, [0]}_{VA+AV}$ is odd.

To complete the proof we notice that under the symmetry transformation
${\rm Ex}(d,s) \times {\cal C}$, the operator ${\cal
Q}^{\pm\,[0]}_{VA+AV}$ is odd, while the operators ${\cal Q}^{\pm
\,[0]}_{SP+PS}$ and ${\cal Q}^{\pm \, [0]}_{T\widetilde{T}}$ are
even. As for the operators ${\cal Q}^{\pm \, [0]}_{VA-AV}$ and ${\cal
Q}^{\pm \,[0]}_{SP-PS}$ they have more complicated transformation
properties, namely ${\cal Q}^{\pm \, [0]}_{VA-AV} \rightarrow {\cal
Q}^{\mp \, [0]}_{VA-AV}$ and ${\cal Q}^{\pm\, [0]}_{SP-PS} \rightarrow
- {\cal Q}^{\mp \, [0]}_{SP-PS}$. The operators that under ${\rm
Ex}(d,s) \times {\cal C}$ have the same transformation properties as
${\cal Q}^{\pm\,[0]}_{VA+AV}$ are thus the combinations ${\cal Q}^{+
\, [0]}_{VA-AV} - {\cal Q}^{- \, [0]}_{VA-AV}$ and ${\cal Q}^{+ \,
[0]}_{SP-PS} + {\cal Q}^{- \, [0]}_{SP-PS}$, but $\SU(4)$ flavour
symmetry prevents ${\cal Q}^{\pm\,[0]}_{VA+AV}$ from mixing with
them. This concludes the proof for ${\cal Q}^{\pm\,[0]}_{VA+AV}$.

\subsection{The operator ${\cal Q}^{\pm \, [1]}_{VA+AV}$}
\label{sec:OP2}

We now consider the case of the operator ${\cal Q}^{\pm \,
[1]}_{VA+AV}$, which differs from the previous one only because the
Wilson parameters of the relevant valence quarks obey the relation
$r_{u'}=-r_d=-r_s=r_{c'}$.

The mixing and renormalization properties of the operator ${\cal
Q}^{\pm\,[1]}_{VA+AV}$ can be brought back to that of the operator
${\cal Q}^{\pm \, [0]}_{VA+AV}$, discussed above, by simply observing
that the change of functional integration variables induced by the
transformation ${\cal R}_{5u'}\times{\cal R}_{5c'}$ does not touch the
expression of ${\cal Q}^{\pm\, [1]}_{VA+AV}$ and leaves invariant the
massless action of the valence quarks, apart from switching the signs
of the Wilson terms of the quark species $u'$ and $c'$. We get in this
way precisely the situation we had before. Since a change of variable
in the functional integral leaves unchanged all the correlators, from
the absence of mixing we proved in the case of ${\cal
Q}^{\pm\,[0]}_{VA+AV}$ we can infer the same result for ${\cal
Q}^{\pm\,[1]}_{VA+AV}$. This ends the proof.

\end{document}